\documentclass[useAME,usenatbib]{mn2e}
\usepackage{graphicx}
\def\lesssim{\mathrel{\hbox{\rlap{\hbox{\lower4pt\hbox{$\sim$}}}\hbox{$<$}}}}
\def\gtrsim{\mathrel{\hbox{\rlap{\hbox{\lower4pt\hbox{$\sim$}}}\hbox{$>$}}}}

\usepackage[T1]{fontenc}
\usepackage{aecompl}
\usepackage{longtable}

\title[Black Holes in Short Period X-ray Binaries]{Black Holes in Short Period X-ray Binaries and the Transition to Radiatively Inefficient Accretion}

\author[G. Knevitt et al. ]{G. Knevitt$^1$, G. A.\ Wynn$^1$, S. Vaughan$^1$ \& M. G. Watson$^1$,
\\
  $^1${Department of Physics \& Astronomy, University of Leicester, 
  Leicester, LE1 7RH, UK}}

\pagerange{\pageref{firstpage}--\pageref{lastpage}} \pubyear{2002}

\def\LaTeX{L\kern-.36em\raise.3ex\hbox{a}\kern-.15em
    T\kern-.1667em\lower.7ex\hbox{E}\kern-.125emX}

\def\gsim{\mathrel{\hbox{\rlap{\hbox{\lower4pt\hbox{$\sim$}}}\hbox{$>$}}}}

\begin{document}

\label{firstpage}
\maketitle

\begin{abstract}
By comparing the orbital period distributions of black hole and neutron star low mass X-ray binaries (LMXBs) in the Ritter-Kolb catalogue \citep{ritter_catalogue_2003} we show that there is statistical evidence for a dearth of black hole systems at short orbital periods ($P_{\rm orb} < 4$ h). This could either be due to a true divergence in orbital period distributions of these two types of system, or to black hole LMXBs being preferentially hidden from view at short orbital periods. We explore the latter possibility, by investigating whether black hole LMXBs could be concealed by a switch to radiatively inefficient accretion at low luminosities. The peak luminosity and the duration of X-ray binary outbursts are related to the disc radius and, hence, the orbital period. At short periods, where the peak outburst luminosity drops close to the threshold for radiatively inefficient accretion, black hole LMXBs have lower outburst luminosities, shorter outburst durations and lower X-ray duty cycles than comparable neutron star systems. These factors can combine to severely reduce the detection probability of short period black hole LMXBs relative to those containing neutron stars. We estimate the outburst properties and orbital period distribution of black hole LMXBs using two models of the transition to radiatively inefficient accretion: an instantaneous drop in accretion efficiency ($\eta$)  to zero, at a fraction ($f$) of the Eddington luminosity ($L_{\rm Edd}$) and a power-law efficiency decrease,  $\eta \propto \dot{M}^n$, for L $< f$ L$_{\rm Edd}$. We show that a population of black hole LMXBs at short orbital periods can only be hidden by a sharp drop in efficiency, either instantaneous or for $n \gsim 3$. This could be achieved by a genuine drop in luminosity or through abrupt spectral changes that shift the accretion power out of a given X-ray band. \end{abstract}


\section{Introduction}

\label{sec:intro}

Low Mass X-ray binaries (LMXBs), comprising a black hole or neutron star primary ($M_1$) and a low mass main sequence or evolved secondary ($M_2 \lesssim 1 M_\odot$), produce X-rays through the accretion of matter onto their primaries  \citep[see reviews by e.g.][]{tanaka_black_1995,van_paradijs_optical_1995,mcclintock_black_2006,done_modelling_2007}.  All black hole and many neutron star LMXBs are transient, experiencing occasional outbursts during which their mass accretion rate and X-ray luminosity increase by several orders of magnitude. These outbursts have typical durations spanning days to months.

The cycle of outburst and quiescence in LMXBs is explained by the disc instability model \citep[see e.g.\ the review by][]{lasota_disc_2001}. Mass transferred from the secondary builds up in an accretion disc around the primary. As the disc density increases, the temperature rises to the point at which hydrogen ionizes. The change in ionization state is associated with a switch to higher viscosity, increasing mass accretion onto the primary and causing an outburst. This enhanced accretion eventually lowers the disc density and temperature to the point at which the viscosity returns to its original value. Unlike cataclysmic variables, where outbursts are terminated by a cooling wave after a few days, the outbursts of LMXBs are prolonged by X-ray irradiation, which heats the disc and traps it in the hot, ionized state \citep{king_light_1998,van_paradijs_accretion_1996}. In sufficiently short period systems the whole disc is irradiated and the majority of the disc mass is accreted. 

The accretion rate during an outburst is known to decay exponentially \citep[e.g.][]{tanaka_x-ray_1996, 1997ApJ...491..312C}, unless the disk is entirely irradiated (see Section 3.2). It was characterised by \citet{king_light_1998} as:
\begin{equation} 
\dot{M} \approx R_D{\nu}{\rho} \: \exp\left(-\frac{3{\nu}t}{R_D^2}\right)  \label{eqn:mdot}.
\end{equation}
\noindent Here $\rho$ is the disc density and $\nu = \alpha_hc_sH$ is the viscosity, where $\alpha_h \sim 0.1$ is the hot state viscosity parameter \citep{1973A&A....24..337S}, $c_s$ is the sound speed, and H is the vertical scale height of the disk. The maximum extent of the disc, $R_D$, is limited by the size of the Roche Lobe of the primary and can be related to the orbital period ($P_{\rm orb}$). For typical mass ratios ($q = M_2/M_1 \le0.1$), $R_D\approx1.77 {\times} 10^{10} m_1^{1/3} P_{\rm orb}^{2/3}({\rm h})$ cm, where $P_{\rm orb}({\rm h})$ is the orbital period in hours and $m_1$ is the primary mass in solar masses.  Both the peak accretion rate ($\dot{M}_{\rm max} \sim R_D{\nu}{\rho}$) and outburst duration ($t_o \sim M_D(R_D)/\dot{M}_{\rm max}$, where $M_D$ is the disc mass; c.f.\ section 3) depend inherently on the orbital period. 

The radiative efficiency of accretion in an LMXB is given by $L = \eta \dot{M}c^2$, where $\eta \sim 0.1$ for a radiatively efficient flow through a thin accretion disc \citep[see e.g, ][]{frank_accretion_2002}. At low accretion rates cooling becomes inefficient and a radiatively inefficient, advection dominated accretion flow (ADAF) can occur \citep{ichimaru_bimodal_1977, rees_ion-supported_1982,narayan_advection-dominated_1995}. The ADAF model is a solution to the hydrodynamical equations of viscous differentially rotating flows with low sub-Eddington accretion rates. The accreting gas has a very low density, leading to an optically thin flow which cannot cool efficiently within an accretion time. The viscous energy is stored in the gas as thermal energy rather than being radiated away, and is advected onto the central compact object. The transition from radiatively efficient to inefficient flow is expected to take place once the accretion luminosity reaches a few percent of the Eddington luminosity ($L_{\rm Edd}$).  For black hole primaries, which lack a hard surface, $\eta \rightarrow 0$ and the accretion energy is carried with the mass flow into the hole or transferred elsewhere, e.g as radio jets or mechanical outflows. Neutron star primaries do not experience a drop in accretion efficiency because the advected energy must always be radiated from the stellar surface. Hence, at sufficiently short periods (and thus small disc radii and peak accretion rates) black hole LMXBs will undergo fainter, shorter outbursts than comparable neutron star systems, making them more difficult to detect.

The relationship between orbital period and peak outburst luminosity for LMXBs is well established \citep[e.g.][]{shahbaz_soft_1998,portegies_zwart_intermediate_2004}. \citet{wu_orbital_2010} studied the outburst luminosities of a sample of transient LMXBs observed by the Rossi X-ray Timing Explorer (\textit{RXTE}). They showed that there was no distinguishable difference in the orbital period-peak outburst luminosity relation of black holes compared to neutron star primaries, when luminosities were measured in Eddington units. However, they suggest that the two populations may diverge at short periods. Along similar lines, \citet{2004A&A...423..321M} note that the low peak outburst luminosities in short period black hole LMXBs can cause them to remain in a low luminosity - hard spectral state, rather than entering the high luminosity - soft state expected for radiatively efficient accretion. This breed of outburst, where sources remain in the low-hard state throughout, has been observed in several sources \citep{2001MNRAS.323..517B, 2004NewA....9..249B}

In addition to establishing a relation between orbital period and peak outburst luminosity, \citet{wu_orbital_2010} point out that the absence, in their dataset, of black holes with orbital periods of $\le 4$hr, may be caused by radiatively inefficient accretion lowering the peak outburst luminosities of these systems. In this paper, we investigate this hypothesis, using the Ritter-Kolb catalogue  \citep{ritter_catalogue_2003} to show that there is statistical evidence for a dearth of black holes in LXMBs at short orbital periods. We suggest that this is caused by the increasing importance of radiatively inefficient accretion in black hole systems lowering not only peak outburst luminosities but also outburst durations and X-ray duty cycles. Additionally, we investigate the nature of the transition to radiatively inefficient accretion, modelling it as an instantaneous change to $\eta$ = 0 at a fraction $f$ of the Eddington luminosity ($L_{\rm Edd}$) and as a power law decrease, $\eta \propto \dot{M}^n$, below $f L_{\rm Edd}$. In Section 2 we present the sample of systems and the statistical evidence for the lack of black hole LMXBs at short orbital periods. In Section 3 we study the effect of a radiative efficiency switch on the peak luminosities and outburst timescales of black hole LMXBs. We determine the conditions under which these effects can hide a short orbital period black hole population in Section 4. This is followed by a discussion of our findings and our conclusions.

\section{Orbital Period Distribution Analysis}

\begin{figure*}
\centering
\includegraphics[width=1.3\textwidth]{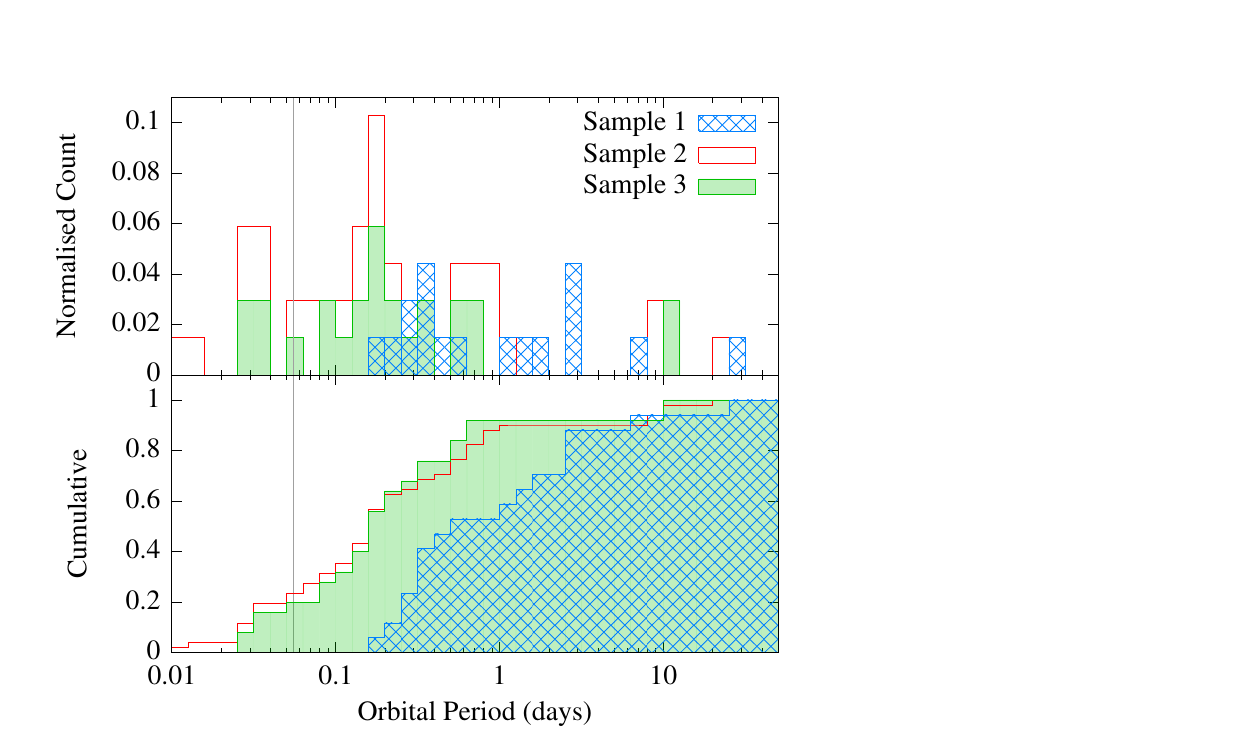}
\caption{Orbital period distributions of neutron star and black hole LMXBs: samples 1, 2 and 3 as described in text. The vertical dashed line separates sources with periods above and below 80 minutes.}
\label{fig:histo}
\end{figure*}

\begin{table*}
 \centering
 \caption{A-D Statistics. In the left hand column, orbital periods of the entire black hole sample (18 sources) is compared to those of all NS LMXBs (51 sources) in row 1, NS LMXBs with orbital periods greater than 80 mins (40 systems) in row 2, all transient NS LMXBs (25 systems) in row 3, and all transient NS LMXBs with orbital periods greater than 80 mins (21 systems) in row 4. The other columns contain the same comparisons but all NS and BH with orbital periods above and below 0.5 days are removed from the middle and right column samples respectively. Results with a $p$-value below 0.01 (high confidence) are highlighted in bold.}
 \begin{tabular}{lccc}
   Black Holes compared to: & A-D Probability (all) & $P_{\rm orb} < 0.5 $ days  &  $P_{\rm orb} > 0.5$ days\\
   \hline
   all NS-LMXBs & $\mathbf{1.2{\times}10^{-3} }$& $\mathbf{8.7{\times} 10^{-4}}$& $0.17$\\
   all NS-LMXBs: $P_{\rm orb}$ $>$ 80 min &  0.016 &$\mathbf{7.2{\times}10^{-3}}$ &0.17 \\
   Transient NS-LMXBs & $\mathbf{1.8{\times}10^{-3}}$ & $\mathbf{5.5{\times}10^{-4}}$& 0.29 \\
   Transient NS-LMXBs: $P_{\rm orb}$ $>$ 80 min  & $\mathbf{9.6{\times}10^{-3}}$ & 0.020 & 0.29\\
   \hline		
 \end{tabular}
   \label{tab:ad}
\end{table*}

If a switch to inefficient accretion in black hole LMXBs at low luminosities reduces their outburst luminosities and timescales, then black holes will be more difficult to detect than neutron stars with comparable accretion rates.  Such an effect may be observed as a divergence in the populations of black hole systems compared to those of neutron stars at short orbital periods. We searched for this within the catalogue of cataclysmic variables, LMXBs and related objects compiled by Ritter and Kolb \citep{ritter_catalogue_2003}, by selecting 3 LMXB samples. 

\begin{itemize}
\item Sample 1: all black hole LMXBs, excluding any unconfirmed candidate systems. (17 systems)

\item Sample 2: all neutron star LMXBs, both transient and persistent, but excluding any globular cluster systems (which are likely to have very different evolutionary histories to Galactic binaries). (51 systems)

\item Sample 3: only the transient neutron star LMXBs from Sample 2 (25 systems)
\end{itemize}

While we see no significant difference in the period distribution of transient and persistent LMXBs (the Anderson-Darling test of the two distributions gives a 0.60 probability that they are the same), we compare both samples 2 and 3 with the black hole population as a control to test for sampling biases caused by the lack of persistent black hole sources.

In all samples, extragalactic sources were removed. In addition, we removed the following systems which are mislabelled in the catalogue: 

\begin{itemize}
\item[(a)] Globular cluster sources (not labeled GC in catalogue):  J1748-2446 \citep{2011AandA...526L...3P}, J1748-2021 \#1 \citep{2008ApJ...674L..45A}, J1748-2021 \#2 \citep{2010ApJ...712L..58A},  NGC 104-X-5  \citep{2002ApJ...564L..17E}, J1623-2631 \citep{2012ApJ...750L...3K}, NGC 104-X7 \citep{2002ApJ...564L..17E}, J1910-5959 \#1  \citep{2009AcA....59..273K} and NGC 104-W37 \citep{2005ApJ...622..556H} ,

\item[(b)] Black hole candidates (not labeled BH? in catalogue): V1408  \citep{2010MNRAS.402.2671R}, V4134 Sgr \citep{2006ApJ...641..410K} J1242+3232 \citep{2007AandA...471L..55C} , J1752-0127 \citep{2009MNRAS.392..309D}
\end{itemize}

In the upper panel of Figure \ref{fig:histo} we plot the normalised histogram of the log[$P_{\rm orb}$] distribution of samples 1, 2 and 3. There is no discernible difference between the two neutron star samples, but the black hole sample shows a distinct lack of sources with orbital periods below $\sim 0.1$ days. There are in fact 25 (12) neutron star LMXBs in sample 2 (3) with orbital periods below the minimum black hole period of 0.17 days (4.1 hr). The cumulative histogram in the bottom panel of Figure \ref{fig:histo} further highlights this difference. 

The vertical dashed line in Figure \ref{fig:histo} separates sources with orbital periods above and below 80 minutes. Below this (conservative) limit, the mass-period relation requires that these systems have either sub-stellar or compact secondaries \citep{king_evolution_1988}. We compare the black hole sample to the NS sample including and excluding sources below this limit, as a check for differences caused by system evolution.

We use the Anderson-Darling test to determine whether the populations of the black hole and neutron star LMXBs are different. The A-D test is similar in purpose to the more familiar Kolmogorov-Smirnov test, and can be used to test whether two samples are consistent with being 
drawn from the same parent distribution. However, the A-D has better 
sensitivity to differences between distributions, especially near their 
tails, compared with the K-S test \citep{feigelson2012}.The A-D test results are shown in Table \ref{tab:ad}. The 4 rows represent comparisons of the black hole sample with (a) sample 2, (b) sample 2 for $P_{\rm orb} > 80 $ min, (c) sample 3 and (d) sample 3 for $P_{\rm orb} > 80 $ min. The A-D $p$-values are shown in the left hand columns; these correspond to the probability of observing a difference as (or more) extreme than actually observed, on the assumption the two parent populations are the same. In addition to the $p$-values for the full samples we show those for sources with orbital periods above and below $P_{\rm orb} = 0.5$ days; this breaks the black hole population roughly in half (8 and 9 sources respectively). Results with $p < 0.01$ are highlighted in bold.

These results suggest that all 4 neutron star samples have different orbital period distributions to the black hole sample across the full period range. This holds when we compare systems with $P_{\rm orb} < 0.5$ days. However, if we remove all sources with periods $<$ 0.5 days, we find no significant difference in the period distributions. We conclude that the black hole and neutron star samples differ at the short end of the orbital period distribution.  

We tested the robustness of this conclusion by adding/removing a few short period systems from each sample. In order to reduce the difference between the samples to the point where the A-D test reports no statistically significant result (i.e. $p>0.05$), we need to remove the eight shortest period NS systems (from the NS transient sample) or include four additional BH systems with periods $<0.1$d. Consequently, the make up of the samples has to change quite considerably in order to influence this result.

\section{Predicted Effects of a Transition to Radiatively Inefficient Accretion}

At high luminosities, a radiative efficiency of $\eta = 0.1$ is a reasonable approximation to black hole and neutron star LMXB outbursts. However,  below 
a fraction $f$ of $L_{\rm Edd}$ \citep[typically a few percent;][]{abramowicz_thermal_1995, 2003A&A...409..697M}, accretion onto black holes becomes radiatively inefficient and $\eta \rightarrow 0$. The efficacy of this switch to radiatively inefficient accretion, in hiding a population of short period black hole LMXBs, depends on the nature of the transition. Here, we investigate two possible forms: 

\begin{description}
\item[A] a sharp, instantaneous switch to $\eta = 0$ at $L \le f L_{Edd}$.
\item[B] A gradual reduction in $\eta$, where below $L \le f L_{Edd}$:
\begin{equation}
{\eta}=0.1 \left(\frac{\dot{M}}{f\dot{M}_{Edd}}\right)^n 
\end{equation}
based on theory by \citet{narayan_advection-dominated_1995}. We use $n = 1$ in subsequent sections, as used in \citet{coriat_revisiting_2012}, but discuss larger values of $n$ in Section 5.2.

\end{description}

A reduction in $\eta$ at low luminosities will reduce the observed duration of an outburst and, in the case of short orbital period systems, may render the entire outburst undetectable.

\begin{figure*}
\centering
\includegraphics[height=0.9\textwidth, angle = -90]{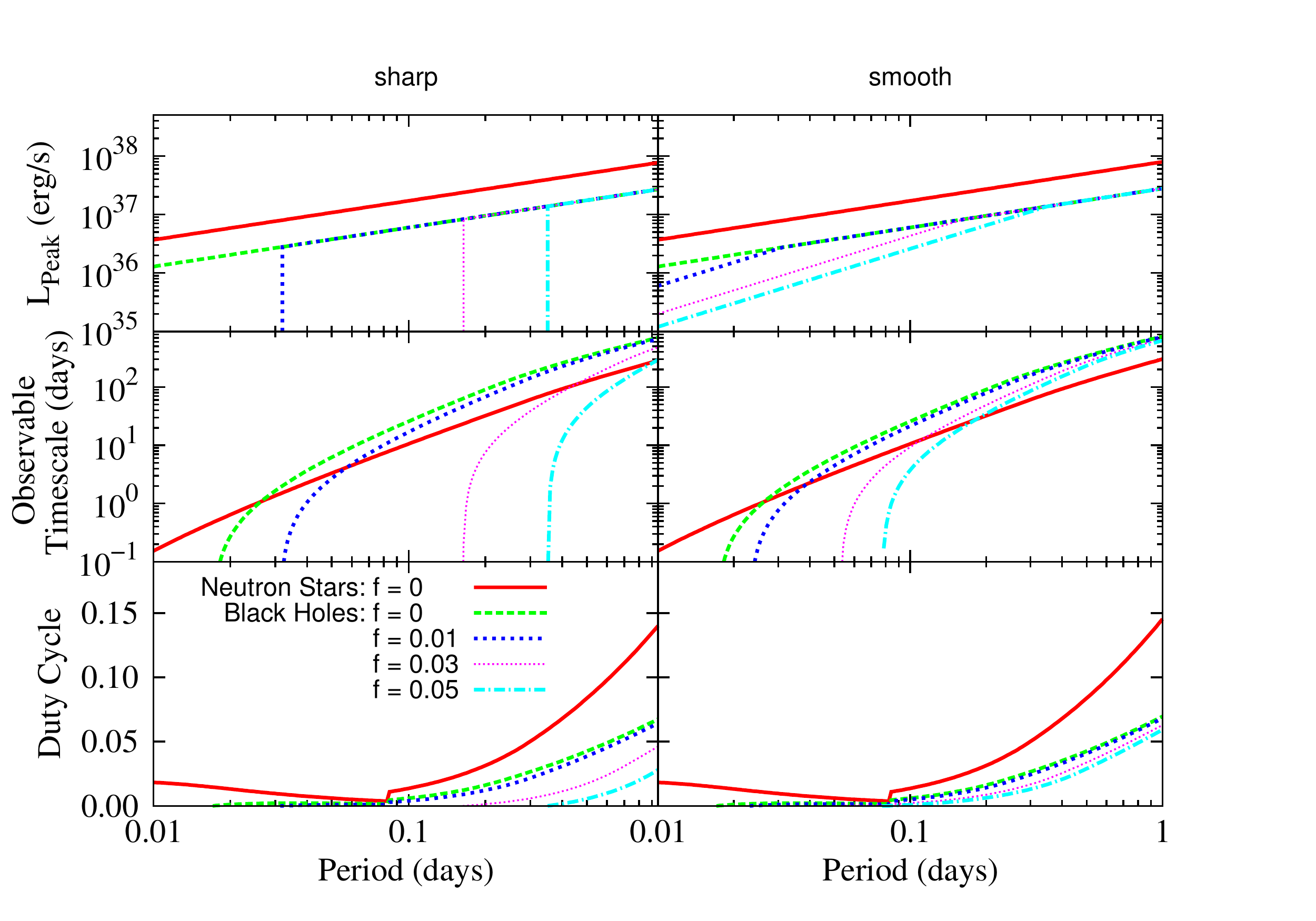}
\caption{Peak 2-10 keV outburst luminosity (upper panels), observable outburst timescale (middle panels) and X-ray duty cycles (lower panels) vs orbital period. The dark blue (thick dotted), purple (fine-dotted) and light blue (dot-dashed) lines show black hole LMXBs with a primary mass of $8 M_{\odot}$ and radiative efficiency switch fractions $f =$ 0.01, 0.03 and 0.05 respectively. Also plotted (solid red line) is a neutron star system with $M_1  = 1.4 M_{\odot}$ and (dashed green line) a $ 8 M_{\odot}$ black hole system with no switch to radiative inefficiency for comparison.  The case A transition (a sharp switch to inefficiency, described in the text) is shown in the left hand panels while the case B (smooth) switch is shown on the right. The observable timescales are defined using a luminosity of L $\sim$$10^{36}$erg/s to signify the "end" of the outburst. This is calculated from the limiting flux, 10 mCrab, with a source distance of 8 kpc. The timescales in the middle panels are replotted for three different distances in  Figure \ref{fig:vary_dist}.  }
\label{fig:varyf}

\end{figure*}
\subsection{Peak Luminosity} 

In the early stages of LMXB outbursts, the accretion rate is known to decay exponentially, following (\ref{eqn:mdot}).  We approximate the disc radius as $R_D \approx 0.7R_{L1}$, where $R_{L1}$ is the Roche radius of the primary:
\begin{equation}
\frac{R_{L1}}{a} =  \frac{0.46 q^{-2/3}}{0.6 q^{-2/3}+ \ln(1+q^{-1/3})}
\end{equation}
\citep{eggleton_approximations_1983}, and $a$ is the binary separation, given by:
\begin{equation} 
a = 3.53 {\times} 10^{10} m_1^{1/3} (1+q)^{1/3} P_{\rm orb} ^{2/3}{\rm (h) \; cm}.
\end{equation}

The secondary mass is fixed at $0.4M_{\odot}$, based on typical values in the Ritter-Kolb catalogue (see Section 4 for further discussion of secondary masses). Following  \citet{king_light_1998}, we take $\rho$ to be $\sim {10^{-8}}$g cm$^{-3}$. They show that $\rho$ is independent of radius, meaning that this value is suitable for short period systems. For neutron stars, we use a value of $2{\times}10^{-8}$ as $\rho$ depends on the surface density (see equation \ref{eqn:sigma}) which is proportional to   $M^{-0.35}$. 

The viscosity is taken as $\nu = \alpha_h c_s H  = \alpha_h c_s (H/R) R_D$ where $\alpha_h \sim$ 0.1, $c_s \simeq 10 \sqrt{(T /10^4{\rm K})}$ km s$^{-1}$ with T, the hydrogen ionisation temperature, $\sim$ 6500K. The alpha prescription above suggests that $\nu$ is linearly proportional to $R_D$. However, observationally, there is significant uncertainty in its value and its dependence on $R_D$. \citet{king_light_1998} choose a value of $\nu = 10^{15}$ cm$^2$s$^{-1}$ for a disc radius of $1{\times}10^{11}$ cm. At the orbital periods $\sim0.1$ days, $R_D$ $\sim 5{\times}10^{10}$ cm. Adopting this as a typical radius, we use  $\nu = 5{\times}10^{14}$ cm$^{2}$ s$^{-1}$. This choice, while consistent with earlier work, also yields peak luminosities in reasonable agreement with the observational results in \citet{wu_orbital_2010}:  $L_{\rm peak} \sim$ (a few $\times 10^{36}$) erg s$^{-1}$ for periods of a few hours. 

At t = 0, the mass accretion rate, $\dot{M}_{\rm peak}=\rho\nu{R_D}$, so the peak luminosity is $L_{\rm bol, peak} = \eta{c^2}{\rho}{\nu}{R_D}$.  Since we wish to compare our estimates to X-ray observations, we convert this bolometric luminosity into the 2-10 keV value, by introducing a correction factor, $f_{\rm corr}$, such that $L_{\rm peak} = L_{\rm bol,peak}/f_{\rm corr}$. For both neutron star and black hole systems, we calculated a simple spectrum, from an accretion disc of specified inner/peak temperature absorbed by a column density of 5$\times$10$^{21}$ $\rm cm^{-2}$; a typical level of Galactic absorption to an X-ray binary. We used this to compute the ratio of bolometric flux (0.002-30 keV) after removing the effect of the absorption, to the flux in the 2-10 keV band including absorption. For this calculation we used XSPEC v12.8 \citep{1996ASPC..101...17A}. In the neutron star case, we used a TBabs absorption model \citep{2000ApJ...542..914W}, whereas we modelled the black hole spectrum with a diskBB model, since, in outburst, its spectrum is dominated by the disc and needs to be treated as a multi-temperature blackbody. A black hole LMXB in outburst has an inner disc temperature of 0.5-1.0 keV, giving $f_{\rm corr} \sim  8.5 - 2.5$.  For neutron star LMXBs, with higher disk temperatures  and thermal emission from their surface and boundary layer, $f_{\rm corr} \sim 1.3-1.2$.  Based on these estimates we use a correction factor of 4 for black holes (corresponding to a temperature of $\sim 0.7$ keV), and 1.3 for neutron stars. 

The uppermost panels of Figure \ref{fig:varyf} show the peak luminosities ($L_{\rm peak}$) as a function of orbital period for black holes of mass $M_1  = 8 M_{\odot}$ using three different values of $f$: 0.01, 0.03 and 0.05, plotted as dark blue dotted, purple dotted and light blue dot-dashed lines respectively. Both a neutron star (solid red line) and a black hole system (green dashed line) with ($f \rightarrow 0$) are shown for comparison on all panels. Our choice of 8$M_{\odot}$ is based on the result of \citet{2010ApJ...725.1918O}, who find a narrow black hole mass distribution at $7.8 \pm 1.2$ $M_{\rm sun}$ \footnote{\citet{2011ApJ...741..103F} found that wider distributions of black hole mass were also consistent with data; this would need to be taken into account if a range of masses were used.} Case A, where $\eta$ drops sharply to 0 at $\dot{M}{\le}f\dot{M}_{\rm Edd}$, is plotted in the left panel. Here,  $f \sim$ 0.03 produces a cut-off in the peak outburst luminosity at orbital periods below $\sim 0.1$ days, matching the observed cut-off of low-orbital period black hole systems in the observational data. The gradient before the cut-off is $2/3$, since $L_{\rm peak} \propto \dot{M}_{\rm max} \propto R_D \propto P_{\rm orb}^{2/3}$. The gradients are consistent with the observed correlation in \citet{wu_orbital_2010}. The different normalisations of the neutron star and black hole $L_{\rm peak}$ - $P_{\rm orb}$ relations are due to our alternate choices of $f_{\rm corr}$ for the two types of system. 

Case B, with a power-law decrease in $\eta$ below $fL_{\rm Edd}$, with $n$  = 1, is plotted in the righthand panel. In this case, the peak luminosity-orbital period relation steepens at short orbital periods when the transition to radiatively inefficient accretion sets in. Here, the factor of $\dot{M}$ in $\eta$ means that $L_{\rm peak} \propto \dot{M}^2 \propto$ $R_D^2 \propto P_{\rm orb}^{4/3}$. While a gradual switch to radiatively inefficient accretion causes the peak luminosities of black hole systems to fall more rapidly at short periods, it does not produce large differences between the two types of system at periods $\sim0.1$ days, unless $f > 0.05$. In this case, the black hole and neutron star gradients diverge below 0.3 days. Observations show a similar gradient and normalisation of peak luminosities down to $\sim0.1$ days, although the predicted effect could be hidden. 

Since our simple analytic models only track emission from the disk, we note that the sharp change in luminosity at $fL_{\rm Edd}$ does not necessarily produce an equally strong observable signature. Sources with peak luminosities below the efficiency threshold are likely to exhibit low-hard outbursts, with power-law spectra that are not dominated by the disk blackbody. In this case, our choice of spectrum, and hence, $f_{\rm corr}$ is no longer valid and it is likely to be larger by factor of $\sim 5$ \citep[see][]{2003A&A...409..697M}. This change in $f_{\rm corr}$ may go some way towards shielding the drop in disk luminosity from the overall bolometric luminosity, which does not change dramatically as the disk moves into a hard state \citep{1997ApJ...477L..95Z,2006MNRAS.367.1113B}

\begin{figure}
\centering
\includegraphics[height=0.9\textwidth, angle = -90]{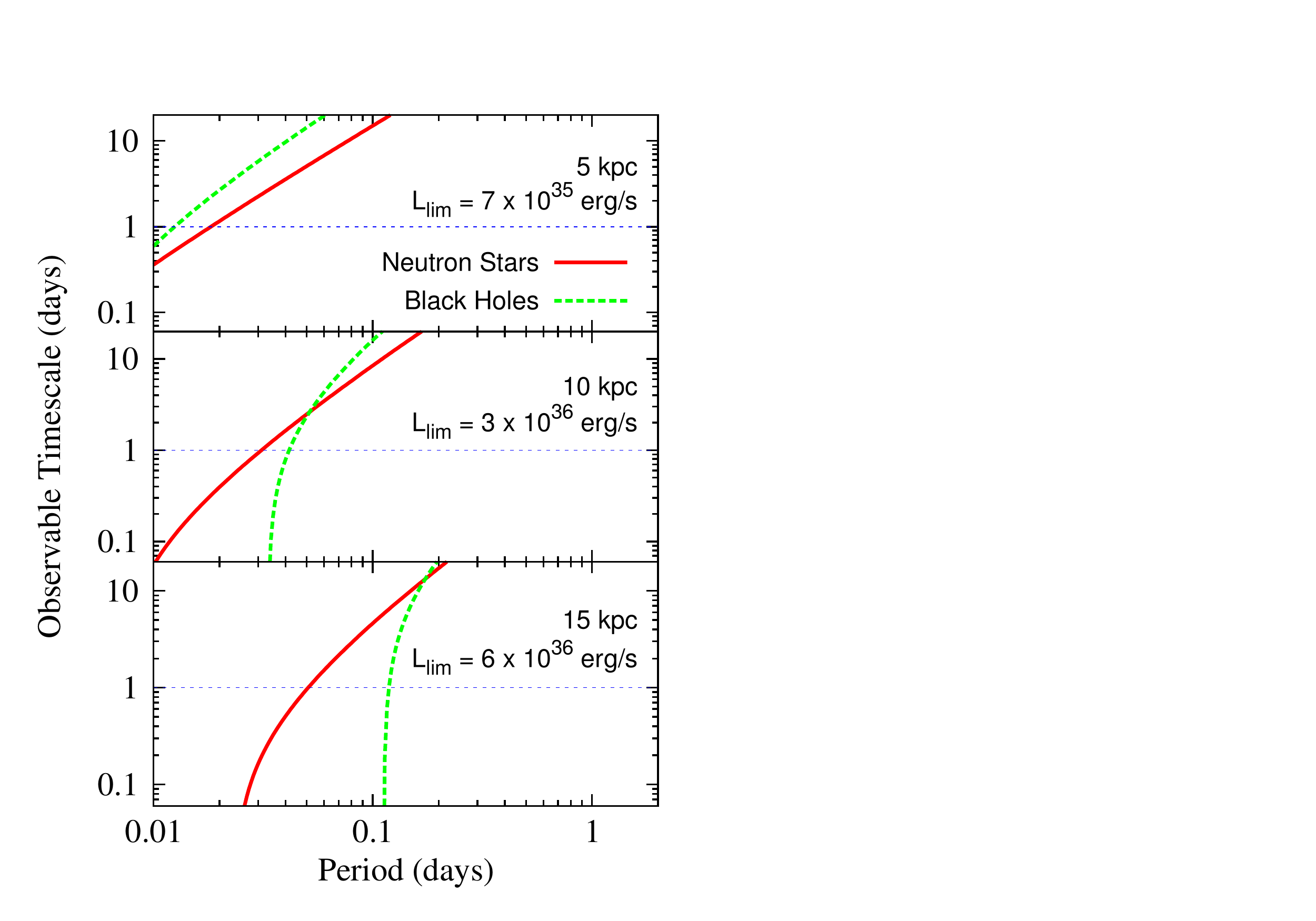}
\caption{Observable timescales ($t_{\rm det}$) are plotted for a black hole LMXB (green dashed line) and a neutron star LMXB (solid red line) as a function of orbital period at three different distances from the source. In the upper panel, the distance is 5 kpc, meaning the outbursts become undetectable at a limiting flux of $7{\times}10^{35}$ erg/s. For the middle panel, at a distance of 10 kpc, the limiting flux is $3{\times}10^{36}$ erg/s, and in the bottom panel the plots are made for a distance of 15 kpc which corresponds to a limiting flux of $6{\times}10^{36}$ erg/s.}
\label{fig:vary_dist}
\end{figure}

\subsection{Outburst Timescales}

From a time $T$ onwards, after the initial exponential fall off, \citet{king_light_1998} show that the mass infall rate follows a linear decay, obeying:

\begin{equation}
 \dot{M} = \left(\frac{3{\nu}}{B_m}\right)^{1/2}\left[M_h^{1/2}(T) -\left(\frac{3{\nu}}{B_n}\right)^{1/2}(t-T)\right]
 \end{equation}
 
\noindent where $T$ is the time taken for the irradiated radius to drop below the disc radius, and is given by:

\begin{equation}
T = \frac{R_D^2}{3{\nu}}\log{\frac{B_n{\nu}{\rho}}{R_D}}
\end{equation}

\noindent and $M_h(T)$ is the irradiated mass at time $T$:

\begin{equation}
M_h(t) = \frac{\rho{R_d}^3}{3}\exp{\frac{-3{\nu}t}{R_D^2}\; .}
\end{equation}

\noindent $B_m$ is defined by $R_h^2  = B_m\dot{M}$ where $R_h$ is the irradiation radius of the disc. The value of $B_m$ depends on $m$; 1 for neutron stars with a hard surface and 2 for black holes without. In our simple calculations we use a value of $B_m = 10^5$ for both systems. We note that the uncertainties in our chosen parameters affect the exact results of our calculations, but not the main conclusions of our study; these are addressed in Section 5.1. 

The outburst ends when $\dot{M} = 0$; with $t_{\rm o} = T + t_{\rm visc}/3$. However, our observations are flux limited, and therefore the outburst appears to end earlier, once its signal drops below a limiting flux, $F_{\rm lim}$.  We define the observable outburst duration, $t_{\rm det}$ to be the time at which the luminosity drops below $L_{\rm lim} = 4{\pi}d^2F_{\rm lim}$. Since the stellar density is highest at the Galactic centre, it is reasonable to assume that this is the location of the highest concentration of LMXBs, so that d $\simeq$ 8kpc.  Using the daily exposure sensitivity of the \textit{RXTE}, we define $F_{\rm lim}$ as $10$ mCrab, and calculate $L_{\rm lim}$ to be $\sim 10^{36}$ $\rm erg s^{-1}$. Our use of the \textit{RXTE} survey to define $F_{\rm lim}$ is explained in Section 4.1. 

In the middle panels of Figure \ref{fig:varyf} we plot the predicted observerable outburst durations. For case A, outbursts are terminated abruptly when the luminosity falls below either $L_{\rm lim}$ or $f L_{\rm Edd}$. As well as the sharp cut-offs corresponding to the peak luminosity graphs in the top panels, we also see the timescales diverging away from the expected duration by an increasing factor as the orbital period drops.  

For case B transitions, a decrease in outburst duration is only evident at orbital periods $\lesssim 0.5$ days. Here a more gradual divergence from the expected outburst timescales occurs. At the minimum observed black hole orbital period of $\sim$ 0.1 days, the expected duration of a black hole outburst is approximately halved for $f= 0.03$. However, it is still longer than the duration of a neutron star outburst, and should not significantly affect detectability.

The relative observable outburst duration of black hole and neutron stars is affected by their distance from us, which determines the minimum luminosity, $L_{\rm lim}$ at which they can be detected. This is illustrated in  Figure \ref{fig:vary_dist}, where the observable timescales ($t_{\rm det}$) are plotted for a black hole LMXB (green dashed line) and a neutron star LMXB (solid red line) as a function of orbital period, at distances of 5, 10 and 15 kpc in the upper, middle and lower panels respectively. These correspond to limiting luminosities of $7{\times}10^{35}$, $3{\times}10^{36}$ and $6{\times}10^{36}$ erg/s respectively. At orbital periods greater than $\sim$ 0.2 days, the outburst duration for black hole systems is longer than that for neutron star systems in all three cases. However, at low orbital periods, neutron star LMXBs may be visible for longer. This is because neutron stars have larger observable peak luminosities (smaller $f_{\rm corr}$). Over time, the neutron star luminosity falls more rapidly than the black hole case (neutron stars have a shorter e-folding times as they have smaller disk radii). If the outburst luminosity of a neutron star LMXB at a given orbital period drops below $L_{\rm lim}$ before it drops below the outburst luminosity of an equivalent black hole system, then the neutron star system will be observed for longer.

\begin{figure}
\includegraphics[height=0.9\textwidth, angle = -90]{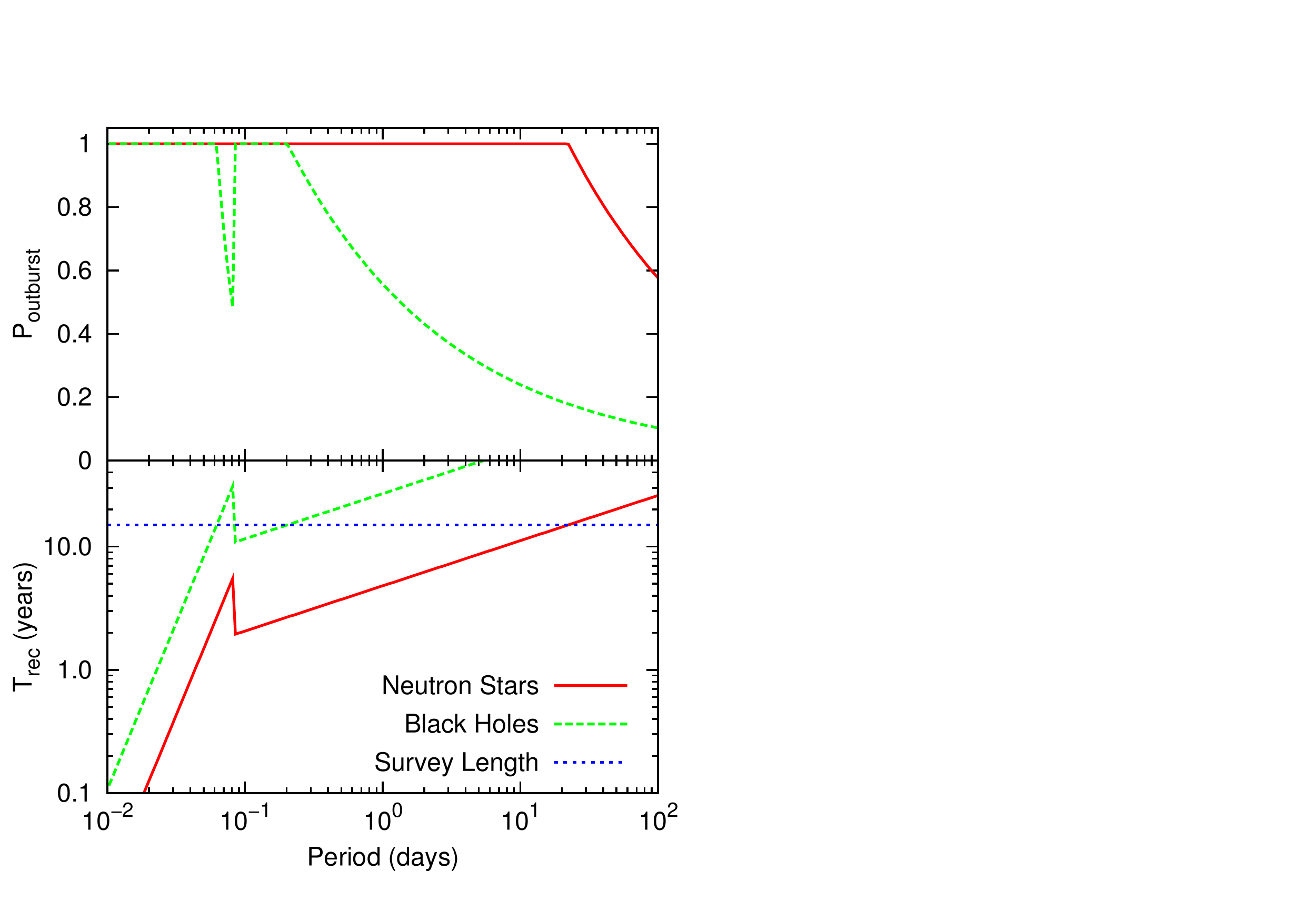}
\caption{Upper panel: Outburst probabilities as a function of orbital period for neutron star and black hole ($M_1$ = 8$M_{\odot}$) LMXBs, plotted as solid red, and green dashed lines respectively. Lower Panel: Outburst recurrence times plotted against orbital period for neutron star and black hole LMXBs. Also shown, as a blue dotted line, is the mission length of the RXTE ASM survey; 15 years.}
\label{fig:P_outburst}
\end{figure}

\subsection{Duty Cycles}
In the lower panels of Figure \ref{fig:varyf} we estimate the outburst duty cycle, which we define as $t_{\rm det} / (t_{q} + t_{o})$, where $t_o$ and $t_q$ are the outburst and quiescent timescales respectively. The quiescent timescale is estimated as $t_q = M_D / |\dot{M}_2|$, where $\dot{M}_2$ is the mass transfer rate from the secondary. We take $M_D$, the disc mass, as:
\begin{equation}
M_D = \int_0^{R_{D}}2{\pi}R_D\Sigma_{\rm max}(R)dR.
\end{equation}

\noindent \citet{cannizzo_outburst_1988} give $\Sigma_{\rm max}$ as:
\begin{equation}
\Sigma_{\rm max} = 11.4 \: \left(\frac{R}{10^{10}{\rm cm}}\right)^{1.05}\left(\frac{M_1}{M_{\odot}}\right)^{-0.35}\alpha^{-0.85}_c {\rm g \: cm}^{-2} \label{eqn:sigma}
\end{equation} 
where $\alpha_c \sim 0.01$ is the cold state viscosity parameter.
Using (\ref{eqn:sigma}) and integrating we obtain:
\begin{equation}
M_D = 2.4{\times}10^{21}\alpha_c^{-0.85}\left(\frac{M_1}{M_{\odot}}\right)^{-0.35}\left(\frac{R_D}{10^{10}{\rm cm}}\right)^{3.05} {\rm g}.
\end{equation}
Using the entire disc mass in the calculation of the quiescent timescale is a good approximation at short orbital periods, where X-ray irradiation is expected to result in the accretion of all or most of the disc. The rate at which the disc is replenished with mass after an outburst is given by $-\dot{M}_2$, which is calculated using the formulae of \citet{king_evolution_1988}:
\begin{equation}
-\dot{M}_2 = \left\{ \begin{array}{ll}\label{eqn:mdot2}
	10^{-10}\left(\frac{P({\rm h})}{2}\right)^{-2/3} M_{\odot}{\rm yr}^{-1} & \mbox{ if $P({\rm h}) < 2$}\\
	6{\times}10^{-10}\left(\frac{P({\rm h})}{3}\right)^{5/3}M_{\odot}{\rm yr}^{-1} & \mbox{otherwise}.
\end{array}
\right.
\end{equation}
The equations above represent mass transfer driven by gravitational radiation ($P({\rm h}) <  2$) and magnetic breaking (otherwise).

It is clear from the lower panels of Figure \ref{fig:varyf} that in case A the X-ray duty cycle of black hole LMXBs $\rightarrow 0$ at periods below 0.1 day for all $f \gtrsim 0.01$. In case B the duty cycle is also noticeably reduced at short orbital periods. In Section 4.3.3., we compare this model to one with a constant $\dot{M}_2$ and find no differences in our conclusions. 


\begin{figure*}
\includegraphics[height=0.9\textwidth, angle = -90]{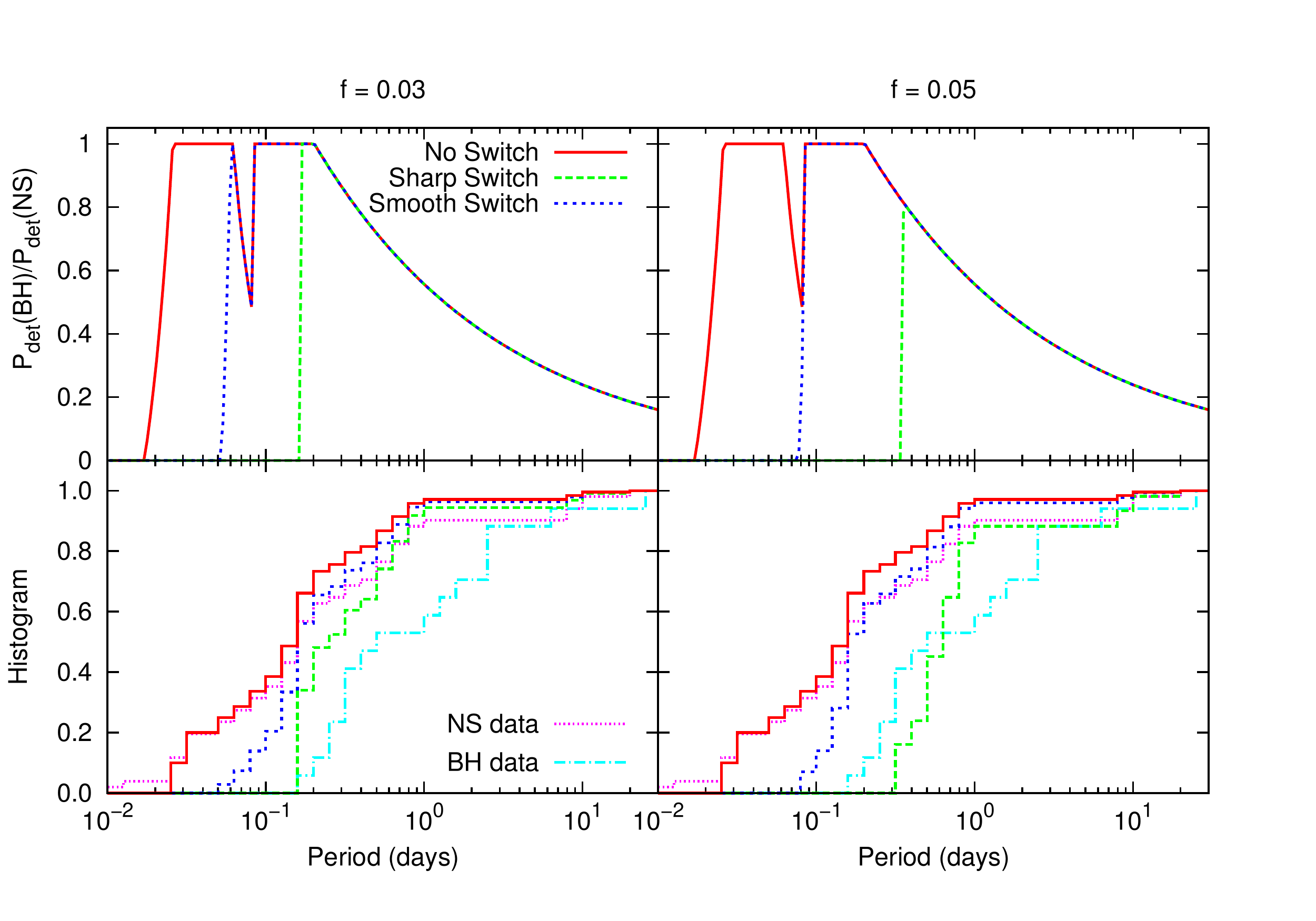}
\caption{Upper panels: the ratio of detection probabilities as a function of orbital period, of black hole ($M_1$ = 8$M_{\odot}$) to neutron star LMXBs, at 8kpc from the Earth. The solid red line shows the ratios with no inefficiency switch during outburst.  The green (dashed) and dark blue (thin dotted) lines include a switch to inefficiency in the black hole systems following case A (sharp switch) and case B (smooth switch) respectively. Results are shown for a switch condition of f = 0.03 in the left hand panels, and f = 0.5 on the right. 
Lower panels:  the predicted orbital period distributions, plotted as normalised cumulative histograms, using the same line styles as the panels above. Also plotted are the observed black hole (light blue dashed line) and neutron star (purple dotted line) distributions.}
\label{fig:nosim}
\end{figure*}

\begin{figure*}
\vspace{-3cm}
\includegraphics[height=0.9\textwidth, angle = -90]{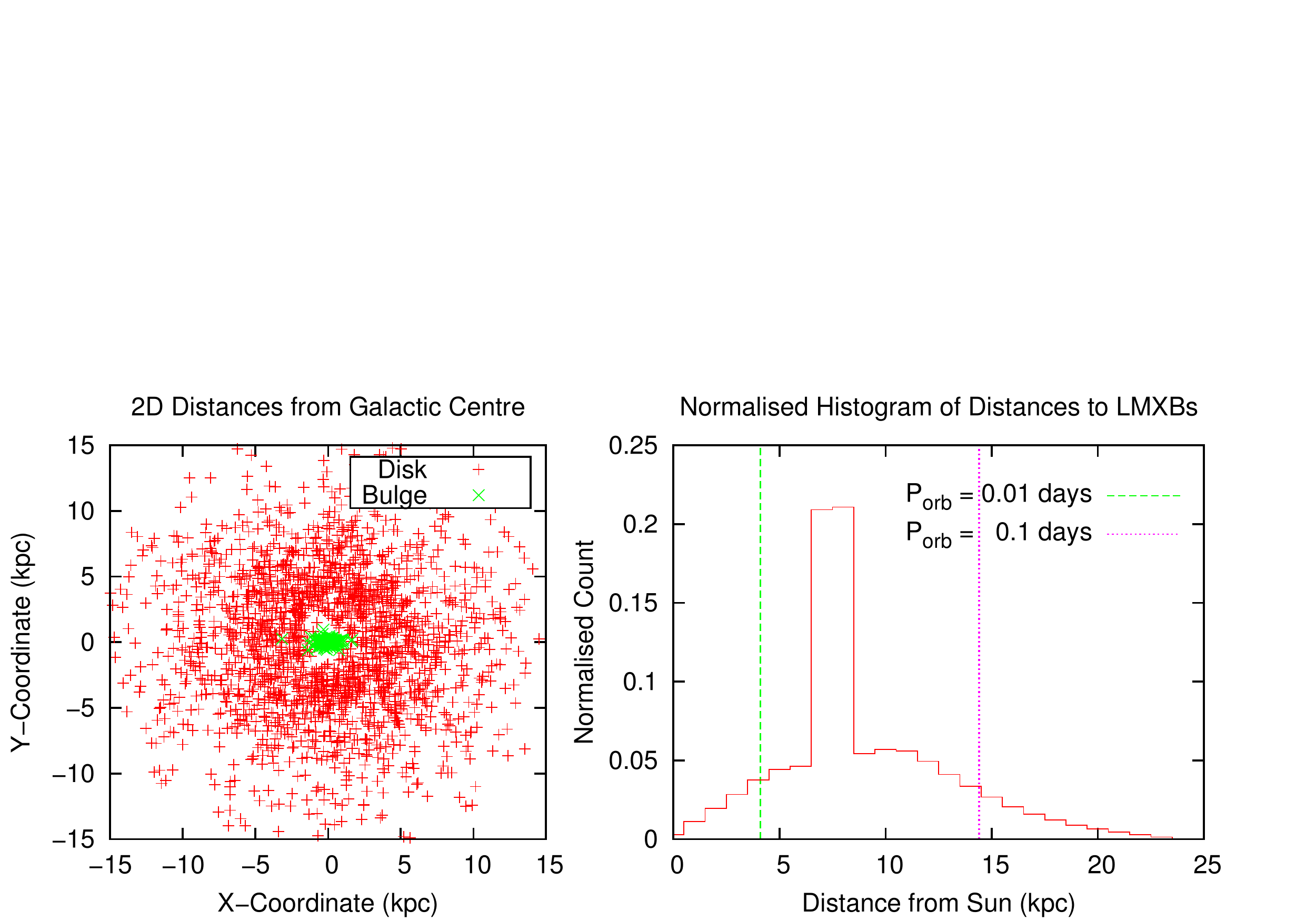}
\caption{Left Panel: 2D positions of 500 LMXBs following the stellar disk distribution (see text), where X and Y are in the plane of the Galaxy. Right Panel: Normalised histogram showing the distances of these LMXBs from the Earth. The green dashed and purpled dotted vertical lines mark the distances out to which black hole LMXBs of orbital period 0.01 and 0.1 days are guaranteed to be observed by the \textit{RXTE} in outburst (i.e. for which $P_{\rm vis} = 1$).}
\label{fig:sim}
\end{figure*}

\begin{figure*}
\includegraphics[height=0.9\textwidth, angle = -90]{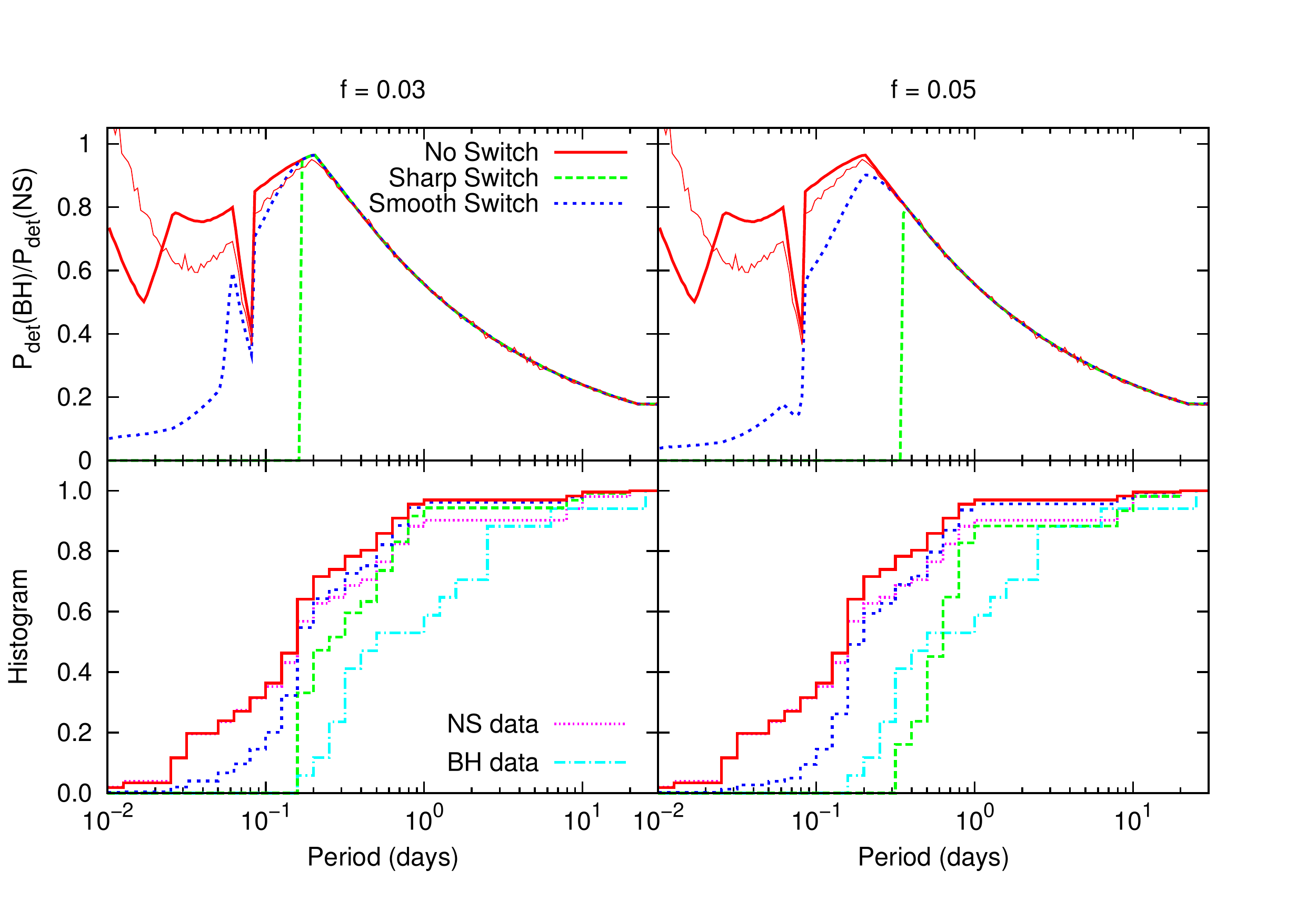}
\caption{Upper panels: the ratio of detection probabilities as a function of orbital period, of black hole ($M_1$ = 8$M_{\odot}$) to neutron star LMXBs, distributed so as to follow the stellar disk population. The solid red line shows the ratios with no inefficiency switch during outburst. The faint red line shows the same, but for a galactic distribution containing only a disk, and no bulge. The green (dashed) and dark blue (thin dotted) lines include a switch to inefficiency in the black hole systems following case A (sharp switch) and case B (smooth switch) respectively. Results are shown for a switch condition of f = 0.03 on the left hand panels, and f = 0.05 on the right. Lower panels: the predicted orbital period distributions, plotted as normalised cumulative histograms, using the same line styles as the panels above. Plotted in light blue dashed line and purple dotted line are the observed black hole and neutron star distributions respectively.
}
\label{fig:MCsim}
\end{figure*}

\begin{figure*}
\includegraphics[height=0.9\textwidth, angle = -90]{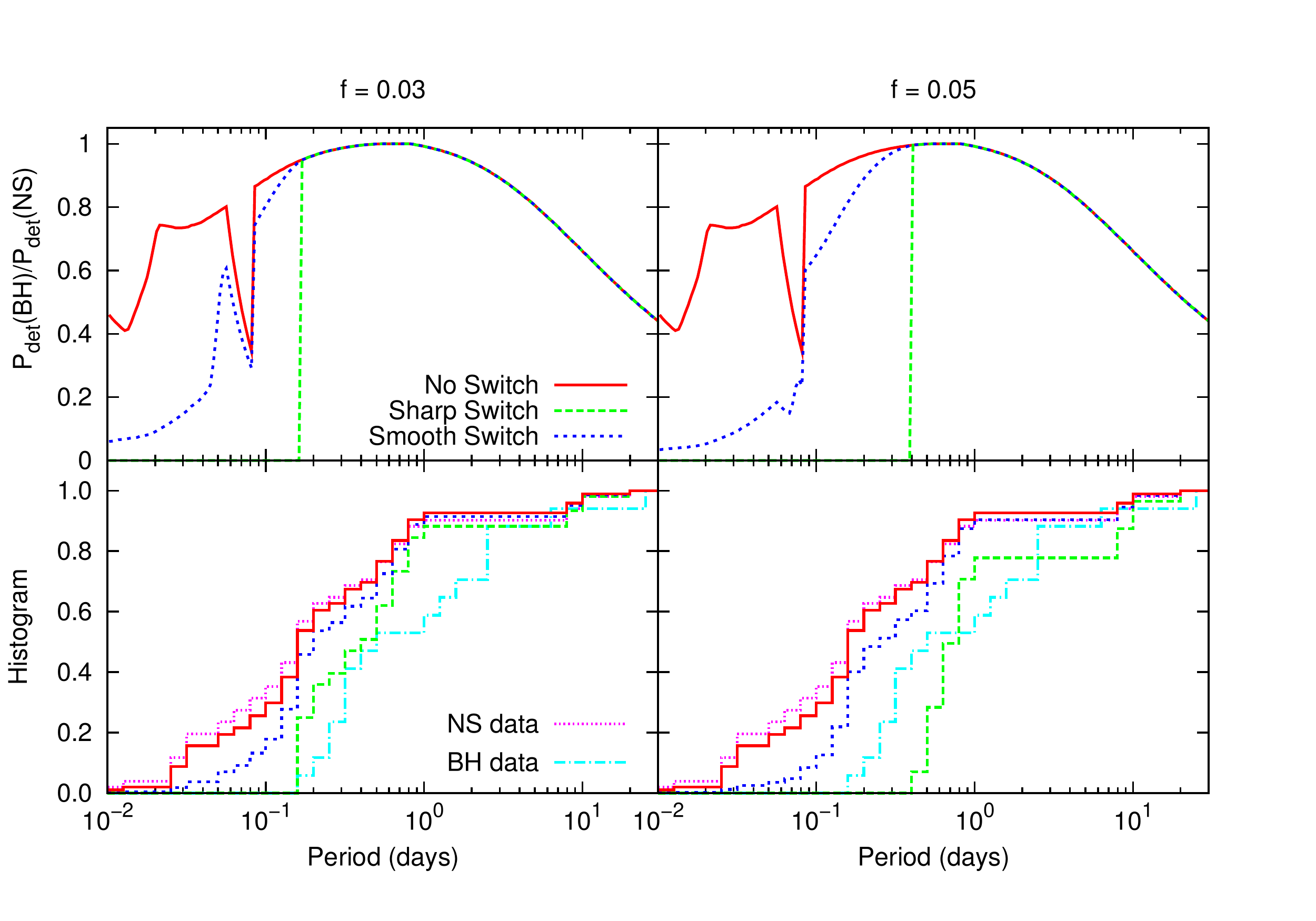}
\caption{All panels plotted as for Figure \ref{fig:MCsim}. The LMXB secondary mass is now following $M_2 = 0.1P_{\rm orb}$}
\label{fig:varM2}
\end{figure*}

\begin{figure*}
\centering
\includegraphics[height=0.7\textwidth, angle = -90]{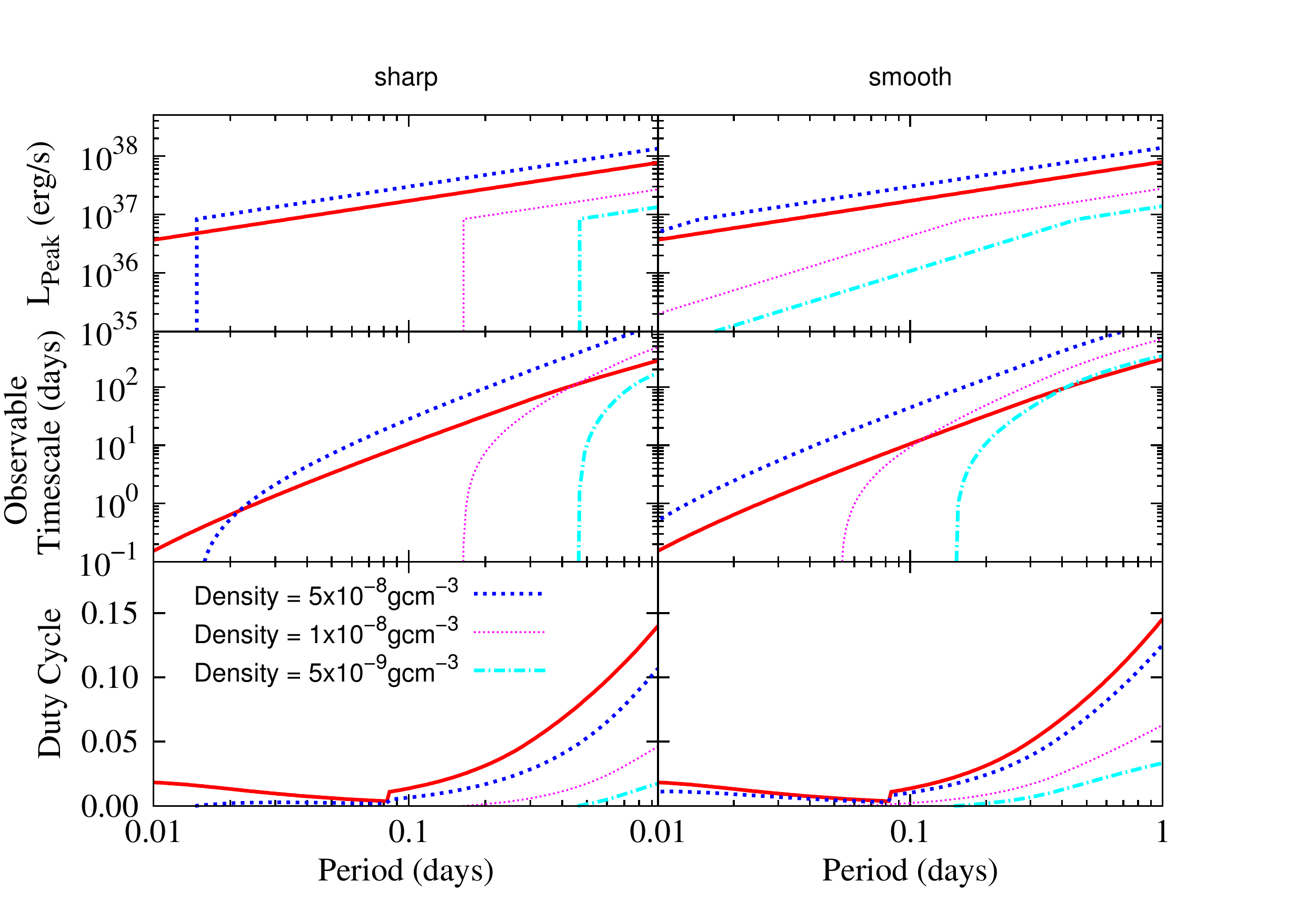}
\caption{The effect of disc density on outburst properties: panels follow those in Figure \ref{fig:varyf}. A black hole of 8 $M_{\odot}$ with no efficiency switch is plotted in solid red. The other 3 lines represent black hole systems ($M_1 = 8 M_{\odot}$) with an efficiency switch at $f = 0.03$. $\rho$ takes values $5{\times}10^{-8}$g cm$^{-3}$ (dark blue dotted line), $1{\times}10^{-8}$g cm$^{-3}$ (purple dotted line, as in Figure 2)  and $5{\times}10^{-9}$g cm$^{-3}$ (light blue dot-dashed line). }
\label{fig:varyrho}
\includegraphics[height=0.7\textwidth, angle = -90]{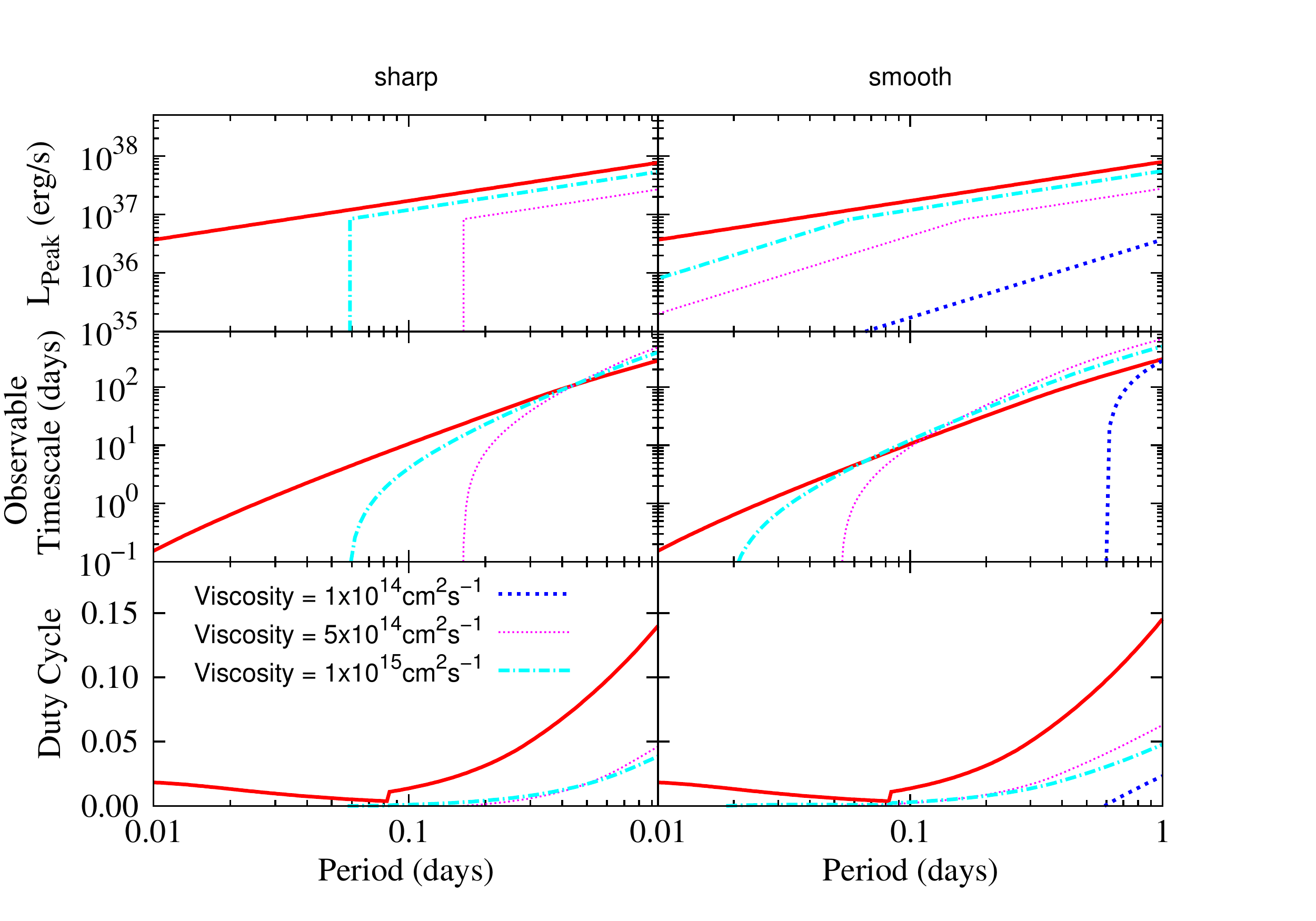}
\caption{The effect of disc viscosity on outburst properties: panels follow those in Figures \ref{fig:varyf} and \ref{fig:varyrho}. A black hole of 8 $M_{\odot}$ with no efficiency switch is plotted in solid red. The other 3 lines represent black hole systems ($M_1 = 8 M_{\odot}$) with an efficiency switch at $f = 0.03$. Here $\nu$ is varied between $1{\times}10^{14}$cm$^2$/s (dark blue dotted line), $5{\times}10^{14}$cm$^2$/s (purple dotted line), and $1{\times}10^{15}$cm$^2$/s (light blue dot-dashed line). }
\label{fig:varynu}
\end{figure*}

\begin{figure*}
\centering
\includegraphics[height=0.7\textwidth, angle = -90]{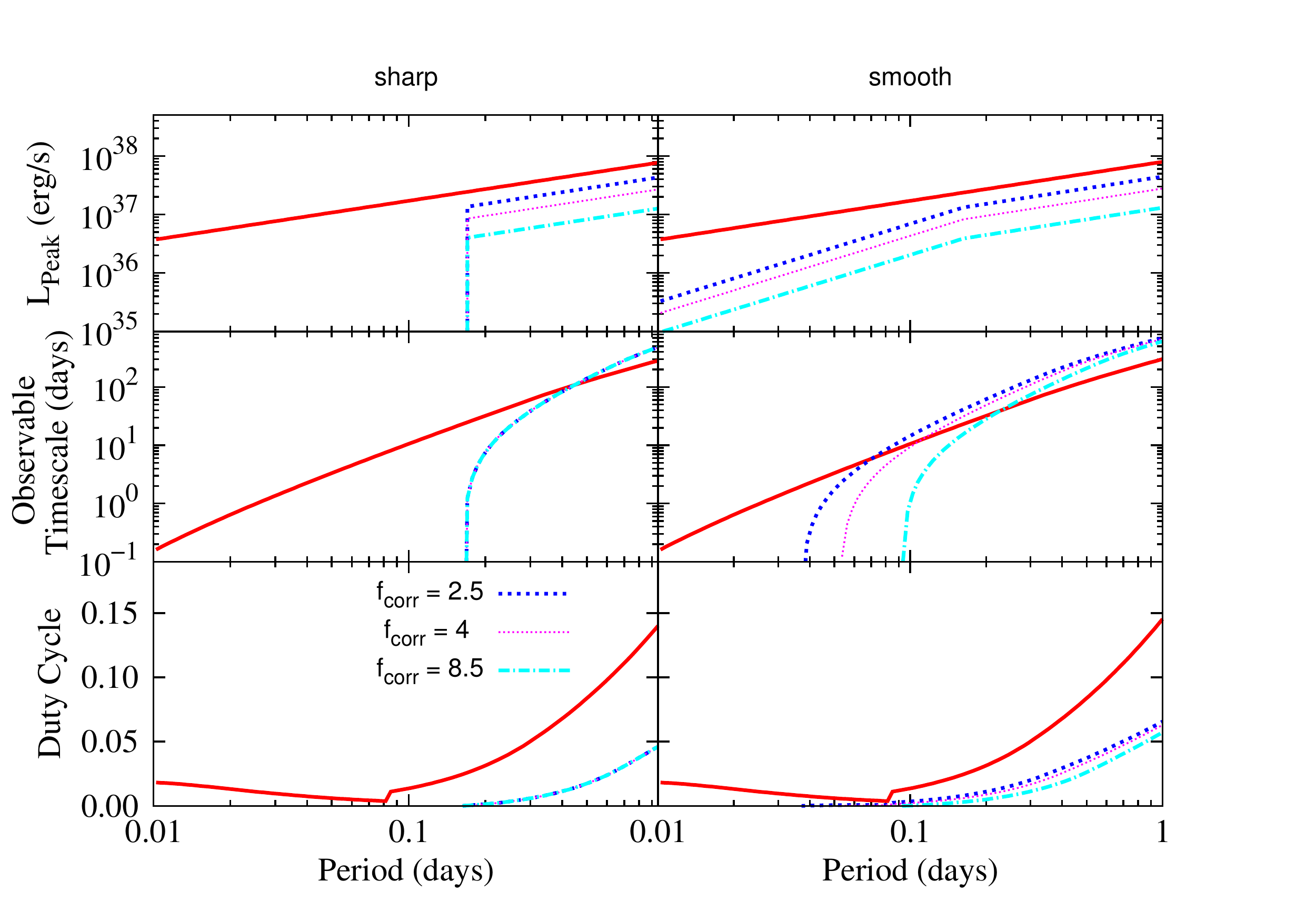}
\caption{The effect of $f_{\rm corr}$ on outburst properties: panels follow those in Figures \ref{fig:varyf} and \ref{fig:varyrho}. A black hole of 8 $M_{\odot}$ with no efficiency switch is plotted in solid red. The other 3 lines represent black hole systems ($M_1 = 8 M_{\odot}$) with an efficiency switch at $f = 0.03$. In this case, $f_{\rm corr}$ takes values of  2.5 (dark blue dotted line), 4 (purple dotted line), and 8.5 (light blue dot-dashed line), which correspond to disc inner temperatures of 1, 0.7 and 0.5 keV respectively. }
\label{fig:varycorr}
\end{figure*}

\begin{figure*}
\includegraphics[height=0.9\textwidth, angle = -90]{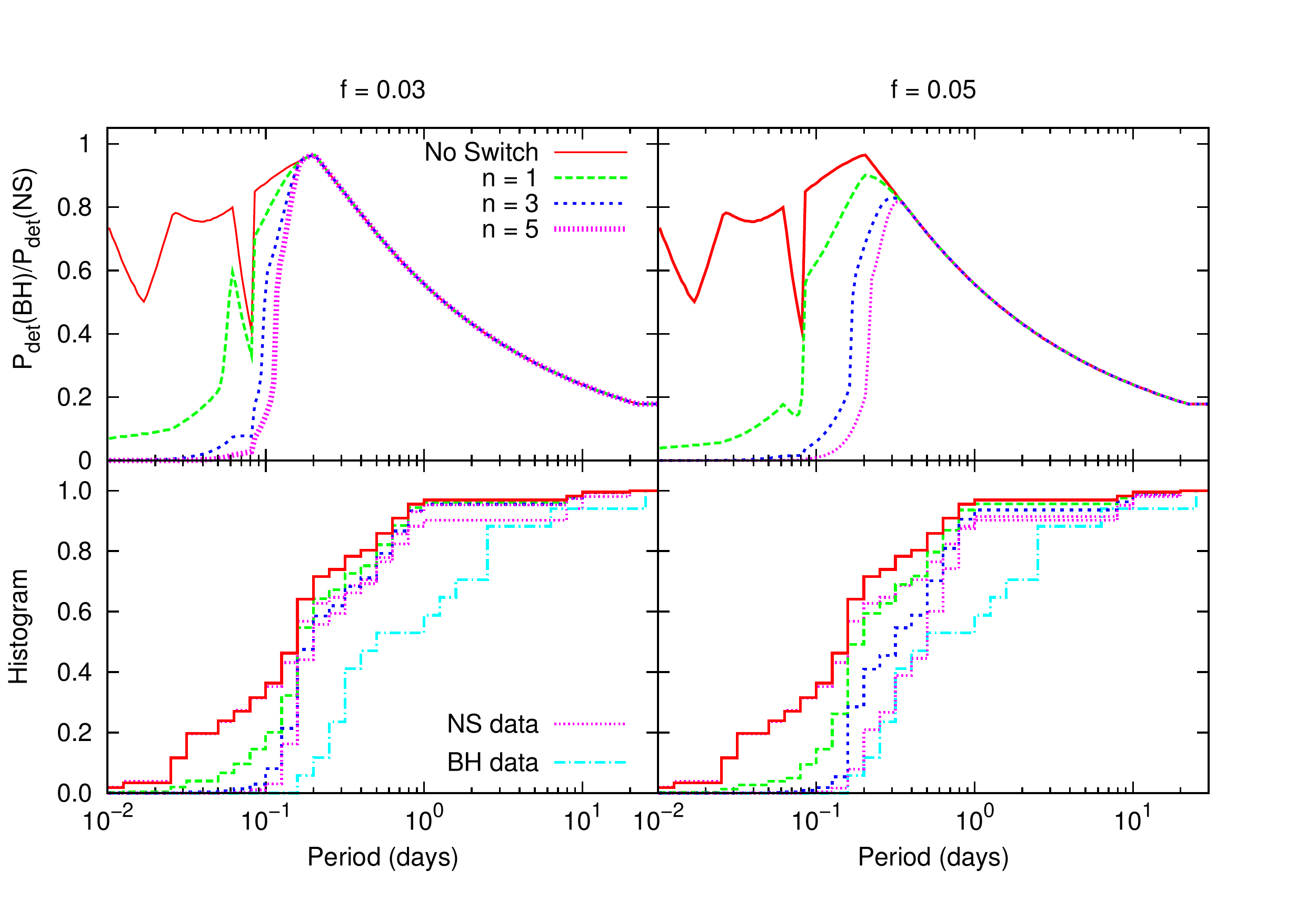}
\caption{Upper panels: the ratio of detection probabilities as a function of orbital period, of black hole ($M_1$ = 8$M_{\odot}$) to neutron star LMXBs. The solid red line shows the ratios with no inefficiency switch during outburst. The other 3 lines (green dashed, dark blue dotted, purple dotted) show the same result with a smooth case B switch and steepness  n = 1, 3, and 5 respectively, as described in the text.  Results are shown for a switch condition of f = 0.03 on the left hand panels, and f = 0.05 on the right. Lower panels: the predicted orbital period distributions, plotted as normalised cumulative histograms, using the same line styles as the panels above. Also plotted are the observed black hole (purple dotted line) and neutron star (light blue dot-dashed line) distributions. 
}
\label{fig:ratios2}
\end{figure*}
 \section{Can a switch to radiatively inefficient accretion explain the observed dearth of BH-LMXBs at short orbital periods?}

We can test the effect of the switch to ADAF in hiding a population of short orbital period black hole systems by predicting the relative detection probabilities of neutron star to black hole LMXBs as a function of orbital period.

\subsection{Observational Requirements}

The All Sky Monitor on board \textit{RXTE} constantly surveyed the sky from March 1996 until January 2012 and discovered many of the known LMXBs in our Galaxy \citep{2006ApJS..163..372W}. The instrument cycled through a series of $\sim 90$ s dwells on different areas of the sky in a stochastic pattern, so that a randomly chosen source was scanned typically 5-10 times per day \citep{1996ApJ...469L..33L}.

The limiting flux, below which a source is undetectable, depends on the instrument used and the exposure time. An all-sky survey is more likely to see a transient object than an instrument with a narrow field of view. For this reason, we use the \textit{RXTE} for all the observational comparisons in this study. The approximate daily sensitivity of the ASM is $10$ mCrab (where $1$ $\rm mCrab$ $\sim2.4{\times}10^{-11}$ $\rm erg\rm s^{-1}\rm cm^{-2}$). We therefore assume an outburst is detectable above $F_{\rm lim}$ = $10$ mCrab, and that the observed outburst duration is $t_{\det} = t(F>F_{\rm lim})$. 

For a transient source to have been detected by the \textit{RXTE}, it must have had at least one outburst during the lifetime of the survey. This depends on the recurrence time $t_{rec}$ =  $t_q$ and the length of the survey ($t_{\rm survey}$ =  15 years), and can be estimated as:

\begin{equation}
P(\ge 1 \; \rm outburst)  = \left\{ \begin{array}{ll}
	t_{\rm rec}/t_{\rm survey} & \mbox{ if $t_{\rm rec} < t_{\rm survey} $}\label{eqn:Pobs}\\
	1 & \mbox{otherwise} 
\end{array}
\right.
\end{equation}

Provided an outburst occurred during the survey time,  it must have been bright enough, and have lasted long enough, for the survey to have detected it. For an outburst to have been unambiguously detected it must have been visible for at least 1 day (private communication, A. Levine $\&$ R. Remillard). If $t_{\rm det} < 1$ day the likelihood of it being observed  ($P_{\rm obs}$) decreases with the decay timescale. We therefore parameterise $P_{\rm obs}$ as:

\begin{equation}
P_{\rm obs}  = \left\{ \begin{array}{ll}
	t_{\rm det}/1 \; {\rm day} & \mbox{ if $t_{\rm det} < 1 \; \rm day $}\label{eqn:Pobs}\\
	1 & \mbox{otherwise} .
\end{array}
\right.
\end{equation}

\noindent The total probability of having an outburst observed by the ASM is therefore: 
\begin{equation}
P_{\rm det} = P_{\rm obs} \; {\times} \; P(\ge 1 \; \rm outburst)
\end{equation}

\noindent and depends on both the mass of the system (i.e. whether it is a black hole or neutron star) and on its orbital period. 

A further constraint on objects forming part of our analysis is whether they have been followed up optically to measure orbital parameters and identify the nature of the primary. We make the assumption here that the fraction of candidate objects which are currently unidentified or poorly studied is independent of the type of primary and its orbital period. In this case, this additional factor will not affect our results.  

\subsection{Relative Detection Probabilities of a Population of LMXBs at 8 kpc}

Initially we make our detection probability calculations for LMXBs located at the Galactic centre, approximately 8 kpc away, as assumed in Section 3.  In this case, the outburst timescales $t_{\det}$ correspond to those in the middle panels of Figure \ref{fig:varyf}. For orbital periods where $t_{\det} > 1$ $ \rm day$, $P_{\rm obs} = 1$ and $P_{\rm det} = P_{\rm outburst}$. In the upper panel of Figure \ref{fig:P_outburst}, outburst probabilities for black hole and neutron star LMXBs are plotted as a function of orbital period, using solid red and green dashed lines respectively. $P_{\rm outburst}$ is a function of $t_{\rm rec}$ which is plotted in the panel below. $P_{\rm outburst}$ drops below 1 only when $t_{\rm rec}$ increases above $t_{\rm survey}$ (blue dashed line). The discontinuity in $t_{\rm rec}$ at 2 hours marks the switch in mass transfer process from magnetic breaking to gravitational radiation in (\ref{eqn:mdot2}). Above this value, $t_{\rm rec} = M_d/|M_2| \propto R_d^3/P_{\rm orb}^{5/3} \propto P_{\rm orb}^{1/3}$, and below,  $t_{\rm rec} \propto R_d^3/P_{\rm orb}^{-2/3} \propto P_{\rm orb}^{8/3}$. Although the discontinuity in $t_{\rm rec}$ produces a negative spike in the black hole outburst probability curve, this feature does not affect our results (see Section 4.3.3 for a discussion.) 

In Figure \ref{fig:nosim} (upper panels) we plot the relative detection probabilities with  $f = 0.03$ (left panel) and $f=0.05$ (right panel). The solid red line shows the ratio of black hole to neutron star detection probabilities without a switch to ADAF, and the green dashed and dark blue dotted lines show the results for sharp and smooth switches (case A and case B) respectively. At large orbital periods, when $P_{\rm obs} = 1$, the ratios plotted are simply the ratios of $P_{\rm outburst}$ shown in the upper panels of Figure \ref{fig:P_outburst}. When there is no efficiency switch, the black hole detection probability drops to zero at $\sim$ 0.01 days, since $t_{\rm det} < 1$ day. When an efficiency switch is included, this drop in detection probability ratios occurs at a higher orbital periods, of $\sim$ 0.05 (0.07) days with a smooth switch and $\sim$ 0.2 (0.3) days with a sharp switch when f = 0.03 (0.05). 

We can now use this detection probability approximation to estimate how the switch to radiatively inefficient accretion affects the orbital period distributions of black hole LMXBs. By convolving the relative detection probability with the observed period distribution of neutron star LMXBs we can predict an expected distribution for black hole systems. We use Sample 2, weighting each neutron star LMXB by the relative detection probability at that orbital period, assuming the population distribution of periods is the same as that observed for neutron stars. These predicted distributions are shown in the bottom panels of Figure \ref{fig:nosim} alongside the observed distributions for black hole and neutron star LMXBs from Figure \ref{fig:histo}. It is clear that the case A efficiency switch is able to reproduce the absence of black hole LMXBs below 0.1 days when f = 0.03, as well as to some extent, predicting the shape of the black hole orbital period distribution. As a simple confirmation, we use these results to estimate the number of black holes expected to have orbital periods below 0.17 days, in a sample the same size as Sample 1 (17 sources). We find it to be zero for a sharp switch, provided f $\ge$ 0.03. It is particularly interesting that this transition lies in the expected range of a few percent of $L_{\rm Edd}$, as any other reason for the relative population difference would give arbitrary values of $f$. Switch B does not produce a significant difference between the black hole and neutron star orbital period distributions, predicting 7 (6) black holes below 0.17 days, for f = 0.03 (0.05). 

\subsection{Relative Detection Probabilities of our Galactic Population of LMXBs}

If all LMXBs are 8 kpc from the Earth, $P_{\rm obs}$ is either 0 or 1, depending on their orbital period. In our Galaxy however, the probability of a source being detected while in outburst depends on its distance, since closer sources will be visible for longer. Therefore, when observing the LMXBs in our Galaxy, $P_{\rm obs}$ may range between 0 and 1, depending on the fraction of sources which are detectable. To address this issue, we now repeat our calculation using a Galactic distribution of LMXBs.

The Milky way stellar population comprises a disk, bulge and spheroidal component. \citet{2002A&A...391..923G} studied the population of LMXBs in our Galaxy, and modelled them using these three components, and found a ratio of 2:1:0.8 of the mass of LMXBs in the disk : bulge : spheroid respectively. The spheroidal component takes account of globular cluster LMXBs, which are not part of our study, so we do not include this in our calculations. We distribute LMXBs in a disk:
\begin{equation}
N_{\rm disk} \left(r,z\right) \propto\exp\left(-a_{\rm disk}\right) \label{eqn:disk_distribution}
\end{equation}
\noindent where
\begin{equation}
a_{\rm disk} = \frac{r_m}{r}+\frac{r}{r_d}+\frac{|z|}{r_z}
\end{equation}
\noindent and a bulge, following the parameterisation of \citet{1997MNRAS.288..365B}
\begin{equation}
N_{\rm bulge} \left(x,y,z\right) \propto \left(1+\frac{a_{\rm bulge}}{a_0}\right)^{-1.8}\exp\left(\frac{a_{\rm bulge}^2}{a_m^2} \right)\label{eqn:bulge_distribution}
\end{equation}
\noindent where
\begin{equation}
a_{\rm bulge} = x^2+\frac{y^2}{\beta^2}+\frac{z^2}{\zeta^2}
\end{equation}
\noindent We use parameters of $r_m$ = 6.5kpc, $r_d$ = 3.5kpc, $r_z$ = 0.41 kpc, $a_m $ = 1.9kpc, $a_0$ = 0.1kpc, $\beta = 0.5$ kpc and $\zeta = $ 0.6 kpc. We normalise both distributions by containing them within 16 kpc, the outer edge of our Galaxy, and by requiring a ratio of 2:1 of binaries in the Galactic disk, and bulge, assuming direct proportionally between number ratios and the mass ratios calculated in \citet{2002A&A...391..923G}

On the left panel of Figure \ref{fig:sim} we plot this distribution, with 500 binaries, where $x$, and $y$ are the cartesian coordinates in the plane of the galaxy, and on the right panel this is converted into a normalised histogram of their distance from the Earth (located at $x = 8\; \rm kpc$, $y = 0\; \rm kpc$) The vertical lines on this plot represent the distances out to which black hole LMXBs with orbital periods of 0.01 (green dashed line) and 0.1 (purple dotted line) days are guaranteed to be observed by \textit{RXTE} when in outburst, i.e. $P_{\rm obs} = 1$. These two lines bracket a significant proportion of the distribution, implying that short orbital period systems in this distance range are unlikely to be observable. In particular it is clear that an efficiency switch, shortening the outburst time, may significantly affect the likelihood of observing short orbital period systems.

\subsubsection{Relative Detection Probabilities}

In order to recreate the detection probabilities and predictions of Figure \ref{fig:nosim} using a distribution in LMXB distances, we run a Monte Carlo simulation of LMXBs in the Galaxy. We create 200 000 LMXBs (100 000 each of black hole and neutron stars) with the same orbital period. These systems are distributed in 3 dimensions over the Galactic disc and bulge by randomly selecting $r$, $\theta$ and $z$ so that they follow the distributions in (\ref{eqn:disk_distribution}) and (\ref{eqn:bulge_distribution}) respectively.  The detection probability is then evaluated for each system and averaged to give an overall detection probability of LMXBs at that orbital period. We run models for orbital periods ranging from 0.01 to 100 days, until a convergence of $0.1\%$ has been reached.

The results are plotted in Figure \ref{fig:MCsim}, in the same style as Figure \ref{fig:nosim}.  Additionally, in the upper panels, we plot the detection probability ratios with no efficiency switch for disk LMXBs only, as a faint red line. This relation differs from the equivalent ratios plotted in Figure \ref{fig:nosim} (solid red line), because $P_{\rm obs}$ now takes fractional values. Below $\sim$ 0.2 days, the most distant black hole LMXBs in the galaxy become undetectable, because they have lower peak luminosities than neutron stars (cf. Figure \ref{fig:varyf}). This causes the black hole LMXB detection probabilities, and hence the relative detection probabilities, to decrease. However, below $\sim$ 0.04 days, the ratios begin to increase again. This is because the neutron star detection probability ratio drops more steeply than the black hole one. This effect is visible in Figure \ref{fig:vary_dist}. From 15kpc (bottom panel), to 5 kpc (top panel), the observable outburst timescale for short orbital period systems increases for both neutron star and black hole LMXBs, as the limiting luminosity becomes lower. However, the rate of increase for black holes is more significant. Therefore, as systems out to greater distances, like 15kpc and 10 kpc become undetectable, the black hole LMXBs become more observable than neutron star systems.  The same shape can be seen in the disk + bulge result (solid red line), but here, another feature is present; a sharp decrease in the relative detection probabilities at an orbital period of $\sim 0.02$ days. This is due to the black hole bulge population becoming undetectable.

While the results for the sharp switch case are the same as in Figure \ref{fig:nosim}, the smooth efficiency switch now alters the predicted black hole orbital period distribution more significantly. However, these results predict 6 (5) black hole systems below 0.17 days when f = 0.03 (0.05), so a smooth switch cannot explain the absence of observed short orbital period black hole LMXBs.  While the sharp switch is still more successful at reproducing the observed distribution, it appears that at higher values of $f$ ($>$ 0.05) the smooth switch may have a similar effect. 

\subsubsection{Varying $M_2$}

In the previous plots, $M_2$ has been fixed at 0.4 $M_{\odot}$. However, the actual dependence of $M_2$ on the systems' other properties is uncertain, and it is likely that $M_2$ increases with orbital period. For this reason, we repeat our models with secondary masses that adhere to the main-sequence, Roche Lobe filling relation,  $M_2 = 0.1 P_{\rm orb}$ \citep{king_evolution_1988}, with the results plotted in Figure \ref{fig:varM2}. Here the relative detection probability curves have a different shape at high orbital periods, leading to minor differences in the predicted black hole population distributions in the lower panels. However, despite this difference, the actual effect of the efficiency switches does not change; a sharp switch, with predictions of 0 (0) observable black hole systems below 0.17 days when f = 0.03 (0.05) is still able to reproduce the divergence in black hole and neutron star systems, whereas a smooth switch, with corresponding predictions of 5 (4) does not. 

\subsubsection{The Discontinuity in $\dot{M_2}$}

The discontinuity (at $P_{\rm orb} \sim 0.1$ days) in the duty cycles plotted in the bottom panels of Figure \ref{fig:varyf}, is due to the switch in accretion mode from magnetic braking to gravitational radiation as described in (\ref{eqn:mdot2}), and also produces a feature in the relative detection probability plots. Since this is in the orbital period range we are interested in, we also checked that our results were not sensitive to this switch by redoing our models with (a) a constant $\dot{M_2}$ and (b) no gravitational radiation mode at short orbital periods. These changes alter P($\ge$ 1 outburst), which describes the shape of the relative detection probability plots. However, they have no effect on $P_{\rm obs}$ which depends on the accretion efficiency switch. Therefore our results, and the effect of different types of accretion switch on the predicted black hole orbital period distributions, are unaltered

\subsubsection{A non-constant Recurrence Time}
Since observations do not show consistent recurrence times for single sources, we have tested the resilience of our results to allowing a range of recurrence times at each orbital period. In the extreme case, the time spent in quiescence after each outburst is independent of any previous event. In this case we can assume a Poisson distribution of outbursts so that: 

\begin{equation}
P(\ge 1 \; \rm outburst) = 1 - e^{-1/\tau} \label{eqn:Pg1}
\end{equation}

\noindent where $1/{\tau}$ = $t_{\rm survey}$/ $t_{\rm rec}$. 

This affects the shape of the black hole to neutrons star detection ratio plots (upper panels of Figure \ref{fig:nosim}), but not enough to change the results in the lower panels, so does not affect our conclusions.

\section{Discussion}

Within the framework of the simple outburst model above, it is clear that the lack of known black hole LMXBs at short orbital periods can be explained by a transition to radiatively inefficient accretion, provided the switch to inefficient accretion is sharp.  We assume a fixed primary mass as well as constant disc densities and viscosities. All of these are likely to vary between observed systems and would produce differences in the details of the orbital period distributions. A more robust treatment aimed at reproducing the observed period distributions would require detailed population synthesis calculations incorporating these factors. Some of the major causes of uncertainty are discussed below.  

\subsection{Choice of Parameters}

The results above are sensitive to our choices of $\rho$, $\nu$, and the black hole $f_{\rm corr}$, which we investigate in Figures \ref{fig:varyrho}, \ref{fig:varynu} and \ref{fig:varycorr}. In Figure \ref{fig:varyrho} we plot the same 6 panels as in Figure \ref{fig:varyf}. The solid red line represents an $8M_{\odot}$ black hole system assuming no switch to inefficiency at low luminosities. The other 3 lines show the effect of varying $\rho$. Each line assumes f = 0.03, and the same $\nu$ as used previously ($5{\times}10^{14}$ cm$^2$ s$^{-1}$). The values of $\rho$ shown are, in order of decreasing density, $5{\times}10^{-8}$g cm$^{-3}$ (dark blue dotted line), $1{\times}10^{-8}$g cm$^{-3}$ (purple, fine dotted line line) and $5{\times}10^{-9}$g cm$^{-3}$ (light blue dot-dashed line). Both the peak luminosity relation and the orbital period at which the sharp switch (switch A) produces a cut-off in the distribution, are affected by changes in density. The smooth switch (case B) also varies. By requiring that the gradient and peak luminosity normalisation match observational expectations it is possible to rule out the higher density cases. 

Similarly in Figure \ref{fig:varynu}, we plot the same systems, with f = 0.03, but this time we keep the density at its original value ($1{\times}10^{-8}$ g cm$^{-3}$) and allow $\nu$ to vary. We choose values of $1{\times}10^{14}$ cm$^{2}$ s$^{-1}$ (dark blue dotted line), $5{\times}10^{14} $cm$^{2}$ s$^{-1}$(purple fine dotted line) and $1{\times}10^{15}$ cm$^{2}$ s$^{-1}$. (light blue dot-dashed line). Varying $\nu$ has a similar effect to varying $\rho$ for the peak luminosities. However, in addition, the timescales, which are inversely proportional to $\nu$,  are extended by decreasing $\nu$. 

In Figure \ref{fig:varycorr},  $\nu$ and $\rho$ are fixed at their original values, and we vary $f_{\rm corr}$, the correction factor applied to the bolometric luminosity of the black hole systems, to estimate the X-ray luminosity. This depends on the inner disc temperature, which is likely to range from 0.5 to 1 keV. We plot values of 2.5 (dark blue dotted line), 4 (purple fine dotted line) and 8.5  (light blue dot-dashed line), which correspond to disk temperatures of 1, 0.7 and 0.5 keV respectively. Varying $f_{\rm corr}$ alters the luminosities of the sharp switch case, and all three panels corresponding to the smooth switch case, in a similar way to varying $\nu$ and $\rho$. Since the inefficiency switch luminosity ($\sim$  2 $\times$ $10^{37}$ ergs s$^{-1}$) is higher than the limiting luminosity at all values of $f_{\rm corr}$ shown, varying this parameter does not affect the timescale or duty cycle plots when a sharp efficiency switch is implemented.

It is clear that our choices of  $\rho$ and $\nu$, which are inherently uncertain, significantly affect the feasibility of the sharp switch to inefficiency explaining a short orbital period black hole deficit. In addition, $f_{\rm corr}$, which depends on the black hole temperature, is likely to affect the normalisation of our results. However, in all cases the smooth switch scenario only produces a "turn-off" the known peak luminosity-orbital period relation at higher orbital periods than observed. 

\subsection{Varying the switch steepness}

We now investigate how sharp the switch needs to be, to produce the observed cut-off. In Figure \ref{fig:ratios2} we vary the dependence of $\eta$ on $\dot{M}$ such that: 

\begin{equation}
{\eta}=0.1 \left(\frac{\dot{M}}{f\dot{M}_{Edd}}\right)^n 
\end{equation}

A larger value of $n$ more closely approximates the sharp switch scenario. Here we show the relative detection probabilities on the top panel for  n = 1 (green dashed),  n = 3 (dark blue dotted) and n = 5 (purple dotted) for f = 0.03 on the left panel and f = 0.05 on the right-hand panel. Indices of $n \gsim 3$ appear to fit the black hole distribution as well as the sharp switch itself, provided f = 0.05, with a prediction of 1(0) observable black hole systems below 0.17 days for n = 3(5). 

\subsection{Observational Issues}

The above discussion is based on detections possible with the \textit{RXTE} ASM, which discovered several new black hole transients during its 15 year mission (e.g. XTE J1118$+$480, XTE J1650-500, XTE J1550$-$56). However, there have been, and still are, many other surveys capable of discovering BHB transients, each with different sensitivity levels, energy coverage, time resolution and sampling. Indeed, \textit{RXTE} also routinely performed scans of the Galactic bulge region with the PCA detectors, capable of detecting fainter sources but reduced angular and temporal coverage compared to the ASM. These scans discovered the BH candidate XTE J1752$-$223 among others.

Presently, the Burst Alert Telescope (BAT) on the $Swift$ mission detects known and new transients in the 15-150 keV range, making it sensitive to the bright low/hard states of black hole X-ray binaries (BHB) during outburst in addition to its higher sensitivity in Crab units. This has been operating since 2004 and discovered several new systems including $Swift$ J1753.5$+$0127, a short period (3.24 hr) black hole candidate \citep{2008ApJ...681.1458Z}. Since 2009 the $MAXI$ monitor has been scanning almost the entire sky in the 2-20 keV X-ray band every 96 minutes and has detected several new transients. Both $Swift$ and $MAXI$ detected a new transient $MAXI$ J1659$+$152, later identified as another short period BHB \citep[2.4 hr; ][]{2013A&A...552A..32K}.

These, together with other past and current missions, increase our sensitivity to BHBs compared to using the \textit{RXTE} ASM alone, increasing the chance of detecting outbursts in Galactic BHBs except at very low peak luminosity and/or very long recurrence times ($>$10 years).

\subsubsection{Comparison to the Ritter-Kolb Catalogue}

We also note that in calculating the relative detection probabilities, we have neglected some of the bias that is likely to exist in the Ritter-Kolb catalogue. Systems with confirmed orbital periods must have had an optical follow-up after their initial X-ray detection. Such a follow-up may be unsuccessful if the X-ray flux is very uncertain, or if the optical/IR counterpart is too faint. Both of these factors are biased against short period systems. We justify neglecting these issues by arguing that they are likely to affect black hole and neutron star systems equally, so would not enter into the ratio calculations.

\subsection{A Hidden Population of Short Orbital Period Black Holes?}

The fact that the values of $f$ fall in the expected theoretical range provides good support for the premise that the transition to radiatively inefficient accretion underpins the lack of black hole LMXBs at short periods. Alternative arguments for the lack of black hole systems (based on system evolution, selection effects etc) would produce arbitrary values of $f$, and the fact that the cut off period for black hole LXMBs happens to agree with theoretical expectations for radiatively inefficient accretion would be purely coincidental.  

Despite this support for a sharp change in $\eta$, observations do not show a sudden change in bolometric luminosity at the efficiency switch boundary.  In addition, observational evidence \citep[see e.g. ][]{1997ApJ...477L..95Z,2005ATel..644....1H} suggests a smooth switch between efficient and inefficient flows. These factors suggest that, rather than being lost to the black hole, accretion power may be shifted out of the X-ray band, through a sharp change in $f_{\rm corr}$. This may mean that systems become detectable in other wavelengths, making it harder for short orbital period black hole LMXBs to "hide" through inefficient accretion. We will consider the detection of hard X-ray dominated LMXB spectra in future work. 

It is worth noting that black hole LMXBs with $P_{\rm orb} \sim 0.1$ day would be expected to produce low luminosity ($\sim$ few \% $L_{\rm Edd}$) and short ($\sim$ 1 day) outbursts. These systems would have low detection probabilities and would be rare, but may be observable in locations where formation conditions were favourable. It may be possible that short period black hole LXMBs form some fraction of the low-luminosity, short-duration very fast X-ray transients observed in the Galactic Centre \citep[e.g.][]{wijnands_rapid_2000, 2013MNRAS.428.1335M}.

\section*{Acknowledgements}
We would like to thank the anonymous referee for their helpful and constructive comments. We are also grateful to Alan Levine and Ron Remillard, for their advice on ASM detections. 
GK acknowledges the support of an STFC studentship. This research is supported by an STFC consolidated grant.

\onecolumn

\begin{longtable}{lcccccp{3.5cm}c}
\caption{List  of LMXBs with Known Orbital Periods}\\
Object Name & RA & Dec & Type & Transient & P$_{\rm orb}$& Source & Sample\\ \hline
V1487 Aql & $19$ $15$ $11.5$& $+10$ $56$ $45$  $1$  & BH  & Yes & 30.8 & \cite{2007ApJ...657..409N}  & (i) \\	
V404 Cyg & $20$ $24$ $03.8$ &  $+33$ $52$ $03$ $1$ &  BH & Yes& 6.4714 & \cite{1994MNRAS.271L...5C} & (i) \\
V4641 Sgr & 18 19 21.6& $-25$ 24 25 1 &BH & Yes  & 2.8173 & \cite{2001ApJ...555..489O} & (i) \\
V1033 Sco & 16 54 00.1 & $-39$ 50 45 1& BH  &Yes & 2.62120 & \cite{2001ApJ...554.1290G} & (i) \\
BW Cir & 13 58 09.9  & $-64$ 44 05 1 & BH & Yes & 2.54451 & \cite{2009ApJS..181..238C} & (i) \\
V821 Ara & 17 02 49.4 & $-48$ 47 23 1  &BH & Yes& 1.7557 & \cite{2003ApJ...583L..95H} & (i) \\
V381 Nor & 15 50 58.8 & $-56$ 28 35 1    & BH & Yes& 1.542033 & \cite{2011ApJ...730...75O} & (i) \\
IL Lup & 15 47 08.3 & $-47$ 40 11 1  & BH & Yes & 1.116407 & \cite{2003IAUS..212..365O} & (i) \\
V2107 Oph & 17 08 14.1& $-25$ 05 32 1& BH & Yes & 0.521 & \cite{1996ApJ...459..226R} & (i) \\
GU Mus & 11 26 26.6 & $-68$ 40 32 1& BH & Yes & 0.432602 &  \cite{1996ApJ...468..380O}& (i) \\
QZ Vul & 20 02 49.4 & $+24$ 14 11 1&BH & Yes& 0.344087 &  \cite{2004AJ....127..481I}& (i) \\
V616 Mon & 06 22 44.4  & $-00$ 20 45 1& BH & Yes & 0.323014 & \cite{2010AandA...516A..58G} & (i) \\
J1650-4957  & 16 50 00.8 & $-49$ 57 45 1 & BH & Yes& 0.3205 &  \cite{2004ApJ...616..376O} & (i)  \\
MM Vel & 10 13 36.3 & $-45$ 04 32 1 & BH  & Yes& 0.285206 & \cite{1999PASP..111..969F}& (i) \\
V406 Vul & 18 58 41.7&  $+22$ 39 30 1 & BH & Yes& 0.274 &\cite{2011MNRAS.413L..15C}  & (i) \\
V518 Per & 04 21 42.8 & $+32$ 54 27 1& BH  & Yes& 0.212160 & \cite{2000MNRAS.317..528W} & (i) \\
KV UMa  & 11 18 10.9& $+48$ 02 13 1  &  BH & Yes & 0.16995 & \cite{2008ApJ...679..732G} & (i) \\ 
J1744-2844 & 17 44 33.1  & $-28$ 44 27.1& NS &  Yes & 11.8367 & \cite{1997IAUC.6530R...1F}& (ii), (iii) \\
0042+3244 & 00 44 48.8  & $+33$ 00 33.2 & NS & Yes  & 11.6 & \cite{1978MNRAS.183P..35W} & (ii),(iii) \\
V1333 Aql & 19 11 15.9& $+00$ 35 06 1 & NS & Yes & 0.78950 & \cite{1998IAUC.6806....2C}& (ii),(iii) \\
J0556-3310 & 05 56 46.3& $-33$ 10 26 1  & NS & Yes & 0.684 & \cite{2012MNRAS.420.3538C}& (ii),(iii) \\
V822 Cen & 14 58 21.9& $-31$ 40 07 1 & NS & Yes & 0.629052 & \cite{2007AandA...470.1033C}& (ii),(iii) \\
QX Nor & 16 12 42.9 & $-52$ 25 23 1  & NS & Yes & 0.5370 & \cite{2002ApJ...568..901W}& (ii),(iii) \\
J1749-2808 & 17 49 31.8& $-28$ 08 05 2 & NS & Yes & 0.367369 & \cite{2010ApJ...717L.149M}& (ii),(iii) \\
J1745-2901 & 17 45 35.7 & $-29$ 01 34 11& NS & Yes & 0.347960 & \cite{2009PASJ...61S..99H}& (ii),(iii) \\
V2134 Oph & 17 02 06.4  & $-29$ 56 44 1 & NS & Yes &  0.296505 &  \cite{2001AandA...376..532O}& (ii),(iii) \\
LZ Aqr & 21 23 14.5  & $-05$ 47 53 1 & NS & Yes & 0.24817 & \cite{2002ApJ...581..570T}& (ii),(iii) \\
V1727 Cyg & 21 31 26.2 & $+47$ 17 25 1 & NS & Yes & 0.218259&\cite{2007AandA...476..301B} & (ii),(iii) \\
AY Sex & 10 23 47.7& $+00$ 38 41 1  & NS & Yes & 0.198096 &  \cite{2009Sci...324.1411A}& (ii),(iii) \\
V5511 Sgr & 18 13 39.0&  $-33$ 46 22 1   & NS & Yes & 0.178110 & \cite{2008MNRAS.391..254C} & (ii),(iii) \\
J1749-2919 & 17 49 55.4 & $-29$ 19 20 1  & NS & Yes & 0.160134 &\cite{2011ATel.3601....1M} & (ii),(iii) \\
UY Vol & 07 48 33.7 & $-67$ 45 08 1 & NS & Yes & 0.159338 &\cite{2009ApJS..183..156W}  & (ii),(iii)\\
J1751-3057 & 17 51 08.7&  $-30$ 57 41 1   & NS & Yes & 0.144531 & \cite{2010MNRAS.407.2575P} & (ii),(iii)\\
J1710-2807 & 17 10 12.5 & $-28$ 07 51 1 & NS & Yes & 0.136711 &  \cite{2011MNRAS.413....2J}& (ii),(iii) \\
V1037 Cas & 00 29 03.1 & $+59$ 34 19 1 & NS & Yes & 0.102362 & \cite{2010ApJ...722..909P} & (ii),(iii) \\
V4634 Sgr & 18 29 28.2 & $-23$ 47 49 2 & NS & Yes & 0.093725 & \cite{2010AstL...36..738M} & (ii),(iii)\\
V4580 Sgr & 18 08 27.6& $-36$ 58 43.1  &  NS & Yes & 0.083902 &  \cite{2008ApJ...675.1468H}& (ii),(iii) \\
J1900-2455 & 19 00 08.7 & $-24$ 55 14 1  & NS & Yes & 0.057815 & \cite{2006ApJ...638..963K} & (ii) , (iii) \\
J1756-2506 & 17 56 57.2 & $-25$ 06 28 1& NS & Yes & 0.037990 & \cite{2007ApJ...668L.147K} & (ii), (iii) \\
J1748-3607 & 17 48 13.1& $-36$ 07 57 1  & NS & Yes & 0.036 &\cite{2012ApJ...753....2G} & (ii),(iii) \\
BW Ant & 09 29 20.2 & $-31$ 23 03 1& NS & Yes & 0.030263 &  \cite{2002ApJ...576L.137G}& (ii),(iii) \\
J1751-3037 & 17 51 13.5& $-30$ 37 23 1   & NS & Yes & 0.029460 & \cite{2002ApJ...575L..21M}  $\;$ $\;$ $\;$& (ii),(iii) \\
V5512 Sgr & 18 14 31.0& $-17$ 09 26 1 & NS & No & 25.07 & \cite{2003ApJ...595.1086C} & (ii) \\
V1341 Cyg & 21 44 41.1 & $+38$ 19 18 1  & NS & No & 9.8445 & \cite{2010MNRAS.401.2517C} & (ii)\\
V395 Car & 09 22 34.7& $-63$ 17 41 1   & NS & No & 9.0026 & \cite{2012MNRAS.424..620A} & (ii)\\
J1538-5542 & 15 38 18.3 & $-55$ 42 12 4  & NS & No & 1.23 &  \cite{2007ATel.1209....1K}& (ii) \\
V1101 Sco & 17 05 44.4 & $-36$ 25 23 1  & NS & No & 0.938 & \cite{1997ApJ...490..401W} & (ii) \\
1624-4904 & 16 28 02.8 & $-49$ 11 55 1 & NS & No & 0.86991 &\cite{2001ApJ...550..962S}  & (ii) \\
1813-1403 & 18 16 01.4 & $-14$ 02 11 1  & NS & No & 0.813 & \cite{1986IAUC.4235....2H} & (ii) \\
V818 Sco & 16 19 55.0& $-15$ 38 25 1  & NS & No & 0.787313 &\cite{1975ApJ...195L..33G}  & (ii) \\
MM Ser & 18 39 57.6 & $+05$ 02 10 1  & NS & No & 0.54 & \cite{1986LNP...266...29M} & (ii) \\
LU TrA & 16 01 02.2 & $-60$ 44 18 1  & NS & No & 0.46229 & \cite{2001AAS...199.0611B} & (ii) \\ 
V691 CrA & 18 25 46.6& $-37$ 06 18 1  & NS & No & 0.232109 & \cite{2010ApJ...709..251B} & (ii) \\
V926 Sco & 17 38 58.1  & $-44$ 27 00 1 & NS & No & 0.193834 & \cite{2006MNRAS.373.1235C} & (ii) \\
V2216 Oph & 17 31 44 .1  & $-16$ 57 41 1 & NS & No & 0.174827 & \cite{2006MNRAS.368..781K} & (ii)\\
GR Mus & 12 57 37.2  & $-69$ 17 19 1& NS & No & 0.163889 & \cite{2009AandA�493..145D} & (ii) \\
V801 Ara & 16 4 55.6 & $-53$ 45 05 1 & NS & No & 0.158047 & \cite{2006MNRAS.373.1235C} & (ii) \\ 
1822-0002 & 18 25 22.0& $-00$ 00 44 1 & NS & No & 0.133 & \cite{2007MNRAS.376.1886S} & (ii) \\
1323-6152 & 13 26 37.0& $-62$ 08 09 1  & Ns & No & 0.122580 & \cite{2006ATel..940....1L} & (ii) \\
UW CrB &16 05 45.7& $+25$ 51 45 1  & NS & no & 0.0771 & \cite{2008ApJ...685..428M} & (ii) \\
J0043+4112 & 00 42 08.7& $+41$ 12 48 1   & NS & No & 0.074 & \cite{2004AandA�419.1045M} & (ii) \\
1705-4402 & 17 08 54.5 & $-44$ 06 07 1 & NS & No & 0.054 & \cite{1987ApJ...323..288L} & (ii) \\
V1405 Aql & 19 18 47.8& $-05$ 14 17 1   & NS & No & 0.034730& \cite{2008ApJ...680.1405H} & (ii) \\
V1055 Ori  & 06 17 07.3 & $+09$ 08 13 1  & NS & No & 0.033713 & \cite{2006MNRAS.370..255N} & (ii) \\
KZ TrA & 16 32 16.7& $-67$ 27 42 1  & NS & No & 0.02876 & \cite{1981ApJ...244.1001M} & (ii) \\
J1806-2924 & 18 06 59.8 & $-29$ 24 30 1& NS & No &  0.027829 & \cite{2007MNRAS.382.1751R} & (ii) \\
QU TrA & 15 47 54.3&  $-62$ 34 11 1 &  NS & No & 0.01264 & \cite{2004ApJ...616L.139W} & (ii) \\
0918-5459 & 09 20 27.0  & $-55$ 12 26 3  & NS & No & 0.01207 & \cite{2011ApJ...729....8Z} & (ii)\\ \hline
\end{longtable}

\end{document}